\documentclass[a4paper,fleqn]{cas-sc}

\usepackage[numbers,sort&compress]{natbib}

\usepackage{amsmath,amssymb,,amsthm,amsfonts,mathtools,bm}

\usepackage{graphicx}
\usepackage{float}
\restylefloat{figure}
\usepackage{booktabs}
\usepackage{multirow}
\usepackage{array}
\usepackage{siunitx}
\usepackage{subcaption}
\usepackage{algorithm}
\usepackage{algpseudocode}
\usepackage{enumitem}
\usepackage{xcolor}
\usepackage{svg}
\usepackage{hyperref}

\usepackage{booktabs}
\usepackage{tabularx}
\usepackage{array}
\usepackage{amssymb}

\theoremstyle{plain}
\newtheorem{lemma}{Lemma}

\theoremstyle{definition}
\newtheorem{property}{Property}
\newtheorem{assumption}{Assumption}
\newtheorem{definition}{Definition}

\newdefinition{remark}{Remark}

\newcommand{\R}{\mathbb{R}}
\newcommand{\norm}[1]{\left\lVert #1 \right\rVert}
\newcommand{\He}{\operatorname{He}}
\newcommand{\lmin}{\lambda_{\min}}
\newcommand{\lmax}{\lambda_{\max}}

\begin{document}

\let\WriteBookmarks\relax
\def\floatpagepagefraction{1}
\def\textpagefraction{.001}

\shorttitle{Adaptive RBFNN Control of Uncertain Bilateral Teleoperation Systems with Delay-Dependent LMI Stability Conditions}
\shortauthors{Ghaemifar et al.}

\title[mode=title]{Adaptive RBFNN Control of Uncertain Bilateral Teleoperation Systems with Delay-Dependent LMI Stability Conditions}

\author[1]{Mohammadali Ghaemifar}
\cormark[1]
\ead{m_ghaemifar@elec.iust.ac.ir}
\credit{Conceptualization, Methodology, Formal analysis, Software, Writing -- original draft}

\author[1]{Arshia Goshtasbi}
\ead{arshia_goshtasbi98@elec.iust.ac.ir}
\credit{Supervision, Validation, Writing -- review and editing}

\author[1]{Arian Hajizadeh}
\ead{arian_hajizadeh@elec.iust.ac.ir}
\credit{Supervision, Validation, Writing -- review and editing}

\author[1]{Armin Attarzadeh}
\ead{armin_attarzadeh@elec.iust.ac.ir}
\credit{Conceptualization, Methodology, Formal analysis, Software, Writing -- original draft}

\author[1]{Erfan Riazati}
\ead{erfan_riazati@elec.iust.ac.ir}
\credit{Conceptualization, Methodology, Formal analysis, Software, Writing -- original draft}

\affiliation[1]{organization={Department of Electrical Engineering, Iran University of Science and Technology (IUST) },
                city={Tehran},
                country={Iran}}

\cortext[cor1]{Corresponding author.}
\begin{abstract}
Bilateral teleoperation requires stability despite uncertain master and slave dynamics and delayed communication channels. Existing radial basis function neural network (RBFNN) controllers mainly differ in uncertainty decomposition, while online adaptive parameters often increase with network size. This paper proposes a compact two-sided adaptive controller for a nonlinear bilateral teleoperator with constant forward and backward delays. Operator and environment impedances are incorporated into the manipulator dynamics, and each side uses only two scalar adaptive estimates: one for the ideal RBF weight norm and another for the combined effects of friction, approximation error, and disturbances. Both estimates are updated through $\sigma$-modification, resulting in only two adaptive parameters regardless of the number of RBF nodes. A Lyapunov–Krasovskii functional integrating sliding-variable energy, estimation errors, and delay-dependent integral terms is developed. Free-weighting matrices based on sliding-surface identities provide delay-dependent matrix conditions that guarantee uniform ultimate boundedness of synchronization errors, sliding variables, and adaptive estimates. Simulations on two 2-degree-of-freedom (2-DOF) revolute manipulators with friction, external disturbance, and stepwise operator inputs demonstrate synchronization within approximately 2 s and reveal the trade-off between accuracy and control chattering.
\end{abstract}



\maketitle


\section{Introduction}

Bilateral teleoperation extends human manipulation capability to locations where direct intervention is difficult, unsafe, or impractical. Manipulator-based teleoperation has consequently been employed in applications ranging from remote maintenance in hazardous facilities to robot-assisted minimally invasive surgery \cite{Rastogi2025,Nasiri2024}. In these systems, commands generated at the master side are transmitted to a remote slave manipulator, and satisfactory operation requires stable and accurate synchronization between the two robots. The continuing development of such systems has made reliable control under realistic communication and modeling conditions an important problem in telerobotics.

Two difficulties are particularly important in this regard. First, communication between the master and slave introduces delay, which may degrade tracking performance and can adversely affect closed-loop stability. The practical influence of communication latency has also been observed experimentally in recent telerobotic systems \cite{Ichihara2025}. Second, an accurate dynamic model is rarely available in practice. Manipulator parameters may be uncertain, while joint friction, approximation errors, external disturbances, and interaction with the operator and environment introduce additional nonlinear effects. A controller intended for practical bilateral operation must therefore provide robustness to both delayed information exchange and uncertain nonlinear dynamics.

This paper considers a nonlinear bilateral teleoperation system with constant communication delays and uncertain master and slave dynamics. The passive components of the operator and environment are represented through mass--spring--damper impedances and incorporated into the corresponding manipulator dynamics, while the remaining nonlinear uncertainty is approximated using radial basis function neural networks (RBFNNs). Rather than adapting every neural-network weight online, each side updates a scalar estimate associated with the norm of the ideal RBFNN weights. A second adaptive quantity estimates a lumped bound containing joint friction, neural approximation residual, and external disturbance. This construction keeps the number of online adaptive parameters independent of the number of RBF nodes while retaining nonlinear approximation capability.

The main objective is to establish the adaptive compensation and the effect of communication delay within the same stability analysis. To this end, a Lyapunov--Krasovskii functional is constructed by combining the sliding-variable energy, adaptive-estimation errors, and delay-dependent integral terms. Free-weighting terms derived from the sliding-variable identities are introduced to obtain feasible delay-dependent stability conditions. The resulting conditions are expressed as linear matrix inequalities (LMIs), whose feasibility can be tested for a prescribed communication delay. Under these conditions, uniform ultimate boundedness of the synchronization errors, sliding variables, and adaptive estimates is established. 

The remainder of this paper is organized as follows. Section~\ref{sec:related_works} reviews the related literature. Section~\ref{sec:preliminaries} introduces the required preliminaries, including the bilateral teleoperator dynamics, operator/environment models, and RBFNN approximation properties. Section~\ref{sec:control_design} develops the proposed controller and adaptive laws. The Lyapunov--Krasovskii stability analysis, the resulting LMI conditions, and the investigation of admissible communication delays are presented in Section~\ref{sec:stability}. Section~\ref{sec:results} presents the simulation results. Finally, Section~\ref{sec:conclusion} concludes the paper.

\section{Related Work}
\label{sec:related_works}

Recent manipulator teleoperation research has considered applications such as robot-assisted minimally invasive surgery \cite{Nasiri2024}, immersive bimanual manipulation \cite{Xu2025}, and force-sensitive robot programming \cite{Sadeghi2026}. From a control perspective, the main developments can be broadly distinguished by how communication delay and uncertain robot dynamics are treated. Early approaches concentrated primarily on preserving stability of the delayed bilateral loop, whereas subsequent adaptive and approximation-based methods placed increasing emphasis on reducing dependence on an accurate dynamic model.

Model uncertainty in a teleoperator is not limited to errors in a few nominal
parameters. Inertia, Coriolis and centrifugal terms, gravity, payload, and
geometric quantities may vary or be imperfectly known, while joint friction
and external disturbances add effects that are difficult to represent
accurately in advance. Communication delay creates a second source of
difficulty and can further deteriorate tracking. Early work on delayed
bilateral teleoperation mainly relied on passivity-based ideas, including
scattering and wave-variable formulations \cite{Anderson1989,Niemeyer1991}.
Lyapunov-based proportional-derivative (PD) \cite{Nuno2008}, sliding-mode \cite{Yang2019}, and robust
$H_\infty$ designs \cite{Sename2007} provided alternative ways of maintaining
stability and tracking performance. These methods remain important baselines,
but model-based designs usually rely on nominal dynamics or known uncertainty
descriptions, whereas passivity-based designs may trade tracking quality for
delay robustness. More importantly, none of these classical approaches is
intended to learn the uncertain nonlinear dynamics online. This motivated the
use of adaptive control in teleoperation.

Adaptive control first addressed uncertainty by exploiting the known structure
of robot dynamics. When the uncertain dynamics can be written in a linearly
parameterized form, a known regressor can be combined with online estimates of
the unknown parameters, as in classical nonlinear adaptive teleoperation
\cite{Nuno2010}. This idea has also been used in position-error-based bilateral
control with constant delay \cite{ID1} and extended to cooperative
single-master/multiple-slave systems with time-varying formation and
communication delay \cite{ID105}. Related robust adaptive and terminal-sliding
formulations have been developed for uncertain manipulators with bounded
disturbances and time-varying delays \cite{ID52}. Such methods are effective
when the uncertain part of the model is described by a known regressor. Their
limitation is equally clear: the structure of the model must still be
specified in advance, and unmodeled nonlinearities or effects outside that
parameterization must be handled through additional bounds or robust terms.
This becomes restrictive when model uncertainty, friction, and external
disturbances have to be treated together. For this reason, fuzzy systems and
neural networks have increasingly been used as online approximators.

Fuzzy logic systems and neural networks relax the need for an explicit linear
parameterization by approximating unknown nonlinear mappings directly.
Adaptive fuzzy synchronization has been studied for teleoperation with
stochastic time-varying delays \cite{Li2011}, while neural controllers have
been combined with prescribed-performance constraints \cite{Yang2015} and
compensation of backlash-like hysteresis \cite{Wang2017}. Fuzzy-neural
backstepping has also been used to learn uncertain robot and environment
dynamics under asymmetric delays \cite{ID74}. A different branch uses wavelet
neural networks: the hybrid neuro-PID scheme in \cite{ID104}, for example,
employs a self-recurrent wavelet identifier and an adaptive proportional
processing network to improve delayed tracking when the slave parameters vary.
These approaches differ considerably in implementation, but most share a
Lyapunov-based adaptation mechanism in which the approximation parameters are
updated together with the controller. Their main practical cost is the number
of online parameters: richer fuzzy rule bases or larger neural representations
generally improve approximation capability at the expense of computation and
tuning. In addition, friction and disturbance effects are often included in a
lumped learned function or handled by auxiliary robust terms. The RBF neural
network is particularly convenient in this setting because it retains
nonlinear approximation capability while remaining linear in the adjustable
weights.

RBFNN-based controllers have therefore received considerable attention in
teleoperation. The single-hidden-layer structure and localized basis functions
allow the unknown nonlinear term to be approximated with a relatively simple
adaptive law. Early RBF position controllers for constant-delay bilateral
manipulators included both acceleration-dependent and acceleration-free
realizations \cite{ID9}. RBF approximation was later combined with
stochastic-delay and LMI analysis \cite{ID34}, and with
wave-variable/passivity mechanisms for time-varying communication channels
\cite{ID14}. Other designs used RBF models together with adaptive robust
control and learned environment representations \cite{ID15}. A closely
related model-approximation approach considered constant delay, uncertain
combined robot/operator/environment dynamics, internal friction, and external
disturbance while reducing the number of online adaptive quantities
\cite{ID36}. Finite-time RBFNN controllers have also been developed for
delayed teleoperation with uncertain robot and interaction dynamics
\cite{ID5}. Taken together, these studies show that RBFNNs provide a useful
compromise between nonlinear approximation and implementable adaptation. At
the same time, the way in which the different uncertainty sources are
represented remains quite different from one design to another.

Several issues are still apparent in the RBFNN literature. The first is the
familiar trade-off between approximation accuracy and online complexity. A
larger set of basis functions can improve approximation over the operating
region, but it also increases the number of parameters that must be updated.
The second issue is how the uncertain terms are grouped before adaptation.
Lumped approximation is common and is not, by itself, a distinguishing
feature. In particular, \cite{ID36} is the closest work to the present problem
setting because it also forms a combined joint-space model and uses an RBF
representation for its uncertain nonlinear dynamics. Its compact adaptive
structure then uses an aggregate robust bound that collects friction,
external disturbance, and approximation-residual effects. In the present
formulation, the nonlinear model uncertainty is likewise treated as a lumped
RBF approximation, whereas the unknown friction and disturbance bounds are
retained as separate adaptive quantities. The relevant distinction is
therefore not ``lumped versus non-lumped'' modeling, but which physical terms
are collected in the neural model and which remain explicit in the adaptive
robust part of the controller. A further issue is the use of acceleration or
delayed derivative information in some RBF designs, although
acceleration-free alternatives have been reported \cite{ID9}. Finally, only a
limited part of this literature first combines the robot and interaction
dynamics in joint space and then applies the neural approximation to the
resulting uncertain model. These differences make the treatment of uncertainty
difficult to compare across existing schemes.

Nevertheless, several issues remain insufficiently addressed in the existing
literature. First, the reviewed RBFNN controllers differ substantially in what
they place inside the lumped neural term and what they leave to robust or
adaptive compensation. This point is especially important for \cite{ID36}:
like the present approach, it uses lumped model approximation under constant
delay and explicitly includes friction and external disturbance. The main
comparison therefore concerns the uncertainty partition. In \cite{ID36},
friction, disturbance, and approximation-residual effects enter an aggregate
bound, whereas the present structure uses the RBFNN for the lumped nonlinear
model uncertainty and retains dedicated adaptive bounds for friction and
external disturbance. Second, the balance between RBF approximation accuracy
and the number of online adaptive parameters is usually fixed through the
chosen network structure rather than addressed explicitly in the control
architecture. Third, unified joint-space modeling of the master/operator and
slave/environment sides before neural approximation is still uncommon,
particularly when the communication delay is included in the stability
analysis. These observations suggest that the remaining problem is not simply
whether uncertainty can be compensated, but how the lumped model approximation
and the remaining uncertainty bounds are organized within one coherent delayed
bilateral control framework.


\newcolumntype{L}[1]{%
  >{\hsize=#1\hsize\linewidth=\hsize\raggedright\arraybackslash}X}
\newcolumntype{C}[1]{%
  >{\hsize=#1\hsize\linewidth=\hsize\centering\arraybackslash}X}

\begin{table}[!t]
\centering
\caption{Comparison of representative adaptive uncertainty-handling methods
for bilateral teleoperation.}
\label{tab:related_comparison}

\footnotesize
\setlength{\tabcolsep}{2.2pt}
\renewcommand{\arraystretch}{1.16}

\begin{tabularx}{\linewidth}{@{}
>{\raggedright\arraybackslash}p{0.62cm}
L{0.90}
L{1.10}
L{1.20}
L{1.40}
C{0.55}
L{0.85}
@{}}

\toprule

\textbf{Ref.} &
\textbf{Method} &
\textbf{Model uncertainty} &
\textbf{Friction / disturbance} &
\textbf{Online adaptation} &
\textbf{Delay} &
\textbf{Stability proof} \\

\midrule

\cite{ID1} &
PEB adaptive (2022) &
Known linear regressor &
Not included in core model &
Physical parameter vector &
Const. &
Lyapunov \\

\cite{ID16} &
Robust adaptive (2020) &
Nominal model + unc. bounds &
Aggregate F/D bounds &
Uncertainty magnitudes &
TV asym. &
Lyapunov / ISS \\

\cite{ID52} &
Adaptive TSM (2021) &
Known dynamic regressor &
Bounded disturbance &
Param. + robust bounds &
TV asym. &
Lyapunov / FT \\

\cite{ID87} &
IT2-FNN adaptive (2020) &
Aggregate fuzzy--NN approx. &
Embedded in aggregate unc. &
FNN params. + learning rate &
TV asym. &
LKF / Lyapunov \\

\cite{ID9} &
Adaptive RBFNN (2013) &
RBF residual approx. &
Friction in residual; D not central &
RBF weights + bound terms &
Const. &
Lyapunov \\

\cite{ID34} &
Stochastic RBFNN (2014) &
RBF model-mismatch approx. &
F/D in residual compensation &
RBF adaptive params. &
Stoch. asym. &
MJ-LKF / LMI \\

\cite{ID5} &
Finite-time RBFNN (2022) &
Composite RBF approx. &
F/D + residual bounds &
Full RBF weights + scalars &
TV asym. &
LKF / FT-UUB \\

\cite{ID36} &
RBFNN uncertainty estimation (2021) &
Reduced-order RBF model approx. &
F + D + NN residual grouped in one bound &
Scalar weight-energy surrogate + aggregate uncertainty estimate &
Const. &
UUB \\

\textbf{Ours} &
\textbf{Adaptive RBFNN estimation} &
\textbf{Joint-space lumped RBF model approx.} &
\textbf{F + D + NN residual handled by adaptive robust bound} &
\textbf{Norm-based RBF adaptation + online residual-bound estimation} &
\textbf{Const.} &
\textbf{LKF / LMI; UUB} \\

\bottomrule

\end{tabularx}

\vspace{1mm}

\begin{minipage}{0.99\linewidth}
\scriptsize
\textit{Abbreviations:}
PEB = position-error-based;
TSM = terminal sliding mode;
IT2-FNN = interval type-2 fuzzy neural network;
F/D = friction/disturbance;
TV = time-varying;
Const. = constant;
Stoch. = stochastic;
MJ = Markov jump;
LKF = Lyapunov--Krasovskii functional;
ISS = input-to-state stability;
FT = finite-time;
UUB = uniform ultimate boundedness.
The proposed method and \cite{ID36} employ the same compact uncertainty-estimation
structure: one scalar estimate associated with the ideal RBF weight norm and one
scalar adaptive bound encompassing friction, external disturbance, and the RBF
approximation residual.
\end{minipage}

\end{table}

The above studies show that adaptive neural approximation and delay compensation have both been extensively investigated in bilateral teleoperation. However, they are commonly introduced through separate elements of the controller and stability analysis. In the present work, the RBFNN approximation, the adaptive compensation of the remaining lumped uncertainty, and the delay-dependent stability conditions are developed within the same closed-loop framework. Based on the preceding discussion, the main contributions of this paper are summarized as follows.

\begin{itemize}

\item A compact two-sided RBFNN adaptive controller is developed for uncertain bilateral teleoperation. For each manipulator, the RBFNN is used to approximate the lumped nonlinear model uncertainty, while only a scalar estimate related to the squared norm of the ideal network weights is updated online. An additional scalar adaptive quantity estimates a bound containing joint friction, the neural approximation residual, and the external disturbance. Consequently, the number of online adaptive parameters does not increase with the number of RBF nodes.

\item Neural adaptation and communication-delay effects are treated within a single Lyapunov--Krasovskii analysis. The proposed functional simultaneously contains the sliding-variable energy, the adaptive-estimation errors, and delay-dependent single- and double-integral terms associated with the communication channels. Thus, the RBFNN compensation and the delay-dependent dissipation are certified in one stability argument rather than assigning the communication delay to an independent passivity transformation, predictor, or delay observer.

\item A free-weighting construction is introduced to obtain feasible delay-dependent stability conditions. A direct bounding procedure leads to an incompatible matrix condition because the delayed boundary terms cannot dominate the resulting current-state terms with positive Krasovskii weighting matrices. To overcome this difficulty, identically zero terms generated from the sliding-variable relations are introduced into the Lyapunov derivative. The resulting free matrices provide additional degrees of freedom and lead to a pair of tractable delay-dependent LMIs.

\item The stability conditions provide a computable admissible-delay criterion. For a prescribed set of controller and adaptation parameters, feasibility of the resulting LMIs can be tested directly as a function of the communication delay. In addition, the closed-loop analysis establishes uniform ultimate boundedness of the sliding variables, synchronization errors, and adaptive estimates. The theoretical results are evaluated through simulations of two nonlinear revolute manipulators subject to model uncertainty, joint friction, external disturbance, and operator excitation.

\end{itemize}
\section{Preliminaries}
\label{sec:preliminaries}

\subsection{Mathematical Preliminaries}
\label{subsec:math_prelim}

Throughout the paper, $\norm{\cdot}$ denotes the Euclidean vector norm and the induced spectral matrix norm, and $\norm{\cdot}_F$ the Frobenius norm. For a symmetric matrix $X$, $X\succ0$ ($X\succeq0$) means that $X$ is positive definite (positive semidefinite), $\lmin\{X\}$ and $\lmax\{X\}$ denote its smallest and largest eigenvalues. The index $i\in\{m,s\}$ refers to the master and the slave robot. The operator and environment damping matrices in the master/slave dynamic model are constant and positive definite, $B_h\succ0$ and $B_e\succ0$; for brevity we set
\begin{equation}
\label{eq:Bpairing}
B_m:=B_h,\qquad B_s:=B_e .
\end{equation}

The master and slave manipulators satisfy the following properties.

\begin{property}
\label{prop:inertia}
For each $i\in\{m,s\}$ the inertia matrix $M_i(q_i)$ is symmetric positive definite and uniformly bounded,
\begin{equation}
\label{eq:inertia-bound}
0<\lmin\{M_i(q_i)\}\,I \;\preceq\; M_i(q_i) \;\preceq\; \lmax\{M_i(q_i)\}\,I<\infty .
\end{equation}
\end{property}

\begin{property}
\label{prop:skew}
Each manipulator model absorbs the constant viscous damping $B_i$ into its velocity-dependent matrix, so that $C_i=C_i^{0}+B_i$, where $C_i^{0}(q_i,\dot q_i)$ is the standard Christoffel matrix for which $\dot M_i-2\,C_i^{0}$ is skew-symmetric. Consequently, for every $\xi\in\R^{n}$,
\begin{equation}
\label{eq:skew}
\xi^{T}\!\left(\dot M_i(q_i)-2\,C_i(q_i,\dot q_i)\right)\xi
=-2\,\xi^{T}B_i\,\xi,
\qquad i\in\{m,s\},
\end{equation}
with $B_i$ defined in \eqref{eq:Bpairing}.
\end{property}

\begin{property}
\label{prop:linear}
There exists a known regressor matrix $Y_z(q_i,\dot q_i,\ddot q_i)\in\R^{n\times r}$ and a constant unknown parameter vector $\theta_z\in\R^{r}$ such that
\begin{equation}
\label{eq:linear}
M_i(q_i)\,\ddot q_i+C_i(q_i,\dot q_i)\,\dot q_i+G_i(q_i)
=Y_z(q_i,\dot q_i,\ddot q_i)\,\theta_z .
\end{equation}
\end{property}

The control design and the subsequent analysis rely on the following assumptions.


\begin{assumption}
\label{ass:disturbance}
For each $i\in\{m,s\}$ there exists an unknown positive constant $d_i$ such that
\begin{equation}
\label{eq:dbound}
\norm{\delta_i(X_i)}+\norm{f_{ci}(\dot q_i)}+\norm{f_i(q_i,\dot q_i)}\le d_i,
\end{equation}
where $f_i$ and $f_{ci}$ denote the internal and external friction terms.
\end{assumption}

\begin{assumption}
\label{ass:delay}
The forward channel (master to slave) and the backward channel (slave to master) introduce constant transmission delays,
\begin{equation}
\label{eq:delay}
\dot T_m=\dot T_s=0,
\qquad
0\le T_m\le\bar T_m,
\qquad
0\le T_s\le\bar T_s,
\end{equation}
where $\bar T_m$ and $\bar T_s$ are known bounds.
\end{assumption}

\begin{definition}[\cite{Khalil2002}]
\label{def:uub}
The solution $x(t)$ is uniformly ultimately bounded with ultimate bound $b$ if there exist positive constants $b$ and $c$, independent of the initial time $t_0$, and, for every $a\in(0,c)$, a time $T=T(a,b)\ge0$ independent of $t_0$, such that
\begin{equation}
\label{eq:uub}
\norm{x(t_0)}\le a
\;\Longrightarrow\;
\norm{x(t)}\le b,
\qquad \forall\,t\ge t_0+T .
\end{equation}
\end{definition}

\begin{lemma}[Jensen's inequality \cite{Gu2003}]
\label{lem:jensen}
For any constant $h>0$, any $R=R^{T}\succ0$, and any integrable $\omega:[t-h,t]\to\R^{n}$,
\begin{equation}
\label{eq:jensen}
\int_{t-h}^{t}\omega(\sigma)^{T}R\,\omega(\sigma)\,d\sigma
\;\ge\;
\frac{1}{h}
\left(\int_{t-h}^{t}\omega(\sigma)\,d\sigma\right)^{\!T}
R
\left(\int_{t-h}^{t}\omega(\sigma)\,d\sigma\right).
\end{equation}
\end{lemma}

\begin{lemma}[Schur complement \cite{Boyd1994}]
\label{lem:schur}
Let $S_{11}=S_{11}^{T}$, $S_{22}=S_{22}^{T}$, and $S_{12}$ be real matrices of compatible dimensions. Then
\begin{equation}
\label{eq:schur}
\begin{bmatrix} S_{11} & S_{12}\\[2pt] S_{12}^{T} & S_{22}\end{bmatrix}\prec0
\quad\Longleftrightarrow\quad
S_{22}\prec0
\ \text{ and }\
S_{11}-S_{12}S_{22}^{-1}S_{12}^{T}\prec0 .
\end{equation}
\end{lemma}

\begin{lemma}[Young's inequality]
\label{lem:young}
For any $a,b\in\R^{n}$ and any scalar $\epsilon>0$,
\begin{equation}
\label{eq:young}
a^{T}b\;\le\;\frac{\epsilon}{2}\,a^{T}a+\frac{1}{2\epsilon}\,b^{T}b .
\end{equation}
\end{lemma}

\subsection{Radial Basis Function Neural Networks}
\label{subsec:rbfnn}

\begin{figure}[H]
    \centering
    \includegraphics[width=0.5\linewidth]{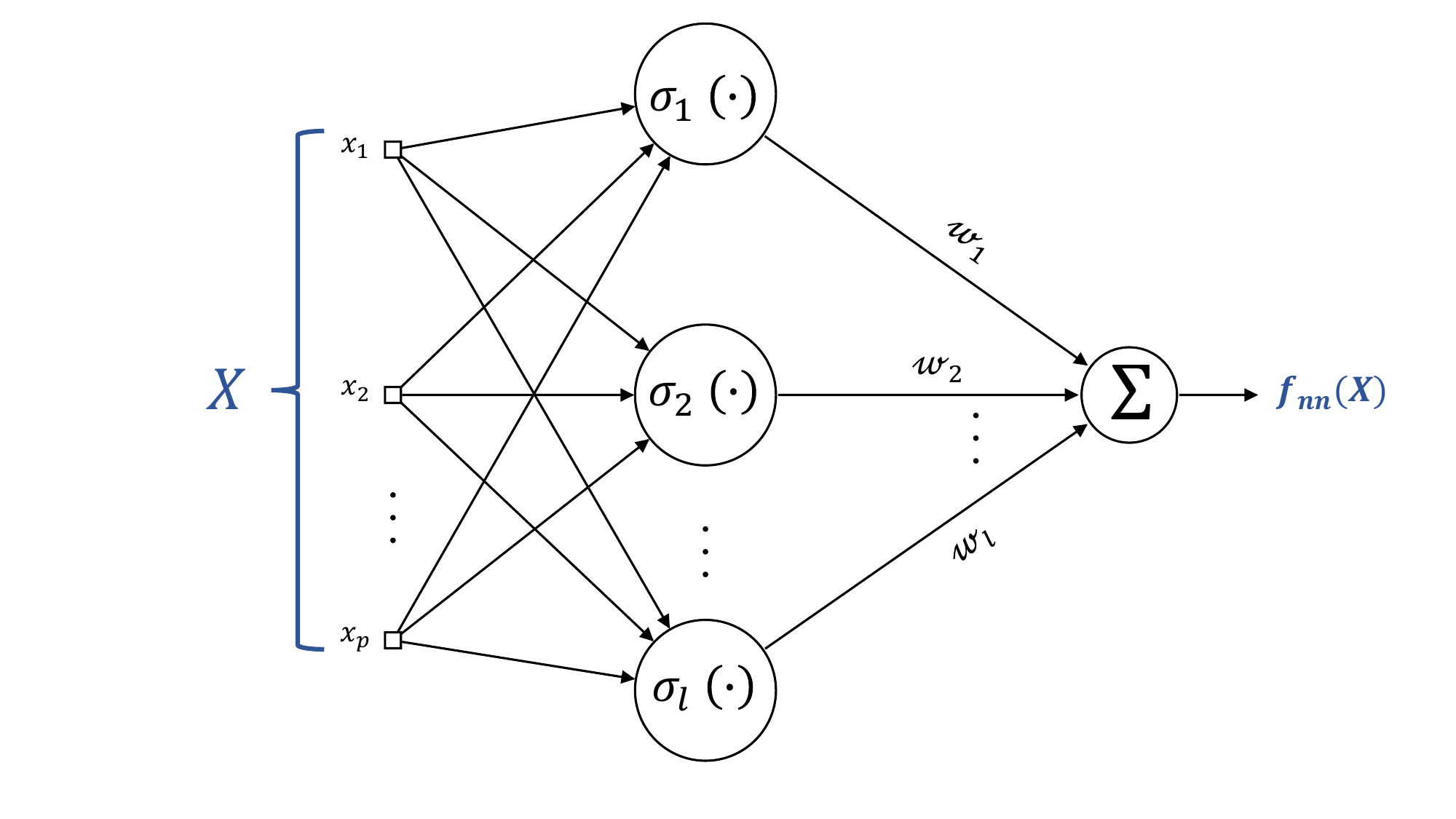}
    \caption{Structure of an RBFNN}
    \label{fig:RBF}
\end{figure}

RBFNNs are known to be approximators for unknown nonlinear functions due to their simple structure and generalization capability. An RBFNN consists of two layers: a hidden layer that maps the input into a high-dimensional space using basis (usually Gaussian) functions, and an output layer that produces a linear combination of the hidden layer outputs with adjustable weights. The structure of an RBFNN can be seen in Figure \ref{fig:RBF}.

For an unknown continuous function $f(X): \mathbb{R}^p \to \mathbb{R}$, the RBFNN approximation is given by
\begin{equation}
f_{nn}(X) = W^T \sigma(X),    
\end{equation}
where $X \in \Omega_X \subset \mathbb{R}^p$ is the input vector, $W = [w_1, w_2, \dots, w_l]^T \in \mathbb{R}^l$ is the weight vector, and $\sigma(X) = [\sigma_1(X), \sigma_2(X), \dots, \sigma_l(X)]^T$ is the vector of Gaussian basis functions. Each Gaussian function is defined as
\begin{equation}
\sigma_j(X) = \exp\left[-\frac{(X - v_j)^T (X - v_j)}{\eta_j^2}\right], \quad j = 1, 2, \dots, l,
\end{equation}
where $v_j$ and $\eta_j > 0$ are the center and width of the $j$-th Gaussian function, respectively.

It has been shown that for a sufficiently large number of nodes $l$, RBFNNs can approximate any continuous function over a compact set $\Omega_X \subset \mathbb{R}^p$ with arbitrary accuracy \cite{Park1991, Haykin1999}. Specifically, there exists an ideal weight vector $W^*$ such that
\begin{equation}
f(X) = W^{*T} \sigma(X) + \varepsilon(X),
\end{equation}
where $\varepsilon(X)$ is the approximation error satisfying $|\varepsilon(X)| \leq \bar{\varepsilon}$, with $\bar{\varepsilon}$ being an unknown positive constant. The ideal weight vector $W^*$ is defined as
\begin{equation}   
W^* := \arg\min_{W \in \mathbb{R}^l} \left\{ \sup_{X \in \Omega_X} \left| F(X) - W^T \sigma(X) \right| \right\}.
\end{equation}

In the teleoperation system, the network input vectors for the master and slave controllers are selected as
\begin{equation}
X_m = \begin{bmatrix} \ddot{q}_s(t - T_s) & \dot{q}_s(t - T_s) & q_s(t - T_s) & \dot{q}_m(t) & q_m(t) \end{bmatrix}^T,
\end{equation}
\begin{equation}
X_s = \begin{bmatrix} \ddot{q}_m(t - T_m) & \dot{q}_m(t - T_m) & q_m(t - T_m) & \dot{q}_s(t) & q_s(t) \end{bmatrix}^T.
\end{equation}
These inputs are chosen to capture the delayed dynamics of the teleoperation system, enabling the RBFNN to approximate the uncertainties and consequently enabling the controller to compensate them.

\subsection{Teleoperation System Dynamics}
\label{subsec:dynamics}

The dynamics of the master--slave robotic system are described by the following Euler--Lagrange equations:
\begin{equation}
\label{eq:maindyn}
\begin{cases}
    M_{q_m}(q_m)\ddot{q}_m + C_{q_m}(q_m, \dot{q}_m)\dot{q}_m + G_{q_m}(q_m) = \tau_m + J_m^T(q_m)F_h, \\
    M_{q_s}(q_s)\ddot{q}_s + C_{q_s}(q_s, \dot{q}_s)\dot{q}_s + G_{q_s}(q_s) = \tau_s - J_s^T(q_s)F_e.
\end{cases}
\end{equation}

For simplicity in this text, $i = m,s$ describes either the master or the slave configuration. $q_i \in \mathbb{R}^n$ represents the joint angular position, and consequently $\dot{q}_i \in \mathbb{R}^n$ and $\ddot{q}_i \in \mathbb{R}^n$ describe the joint angular velocity and acceleration. $M_{q_i}(q_i) \in \mathbb{R}^{n \times n}$ is the inertia matrix, which, as stated in Property~\ref{prop:inertia}, is symmetric positive definite and uniformly bounded. $C_{q_i}(q_i, \dot{q}_i) \in \mathbb{R}^{n \times n}$ denotes the Coriolis and centripetal matrix; by Property~\ref{prop:skew}, this matrix incorporates the constant viscous damping $B_i$ so that $\dot{M}_{q_i} - 2C_{q_i}$ satisfies the skew-symmetry relation given in~\eqref{eq:skew}. $G_{q_i}(q_i) \in \mathbb{R}^n$ represents the generalized gravity vector. $\tau_i \in \mathbb{R}^n$ is the actuator torque, selected as the control input. $J_i \in \mathbb{R}^{n\times p}$ denotes the Jacobian matrix that maps the human operator or environmental forces applied to the end-effector ($F_k \in \mathbb{R}^p$, $k = h,e$) to the robot joint torque, and $p$ is the dimension of the task space.

The operator and environment dynamics are described by the following linear time-invariant (LTI) models:
\begin{equation}
\begin{cases} 
    F_h = f_h^* - M_h \ddot{x}_m - B_h \dot{x}_m - K_h x_m \\ 
    F_e = f_e^* + M_e \ddot{x}_s + B_e \dot{x}_s + K_e x_s 
\end{cases}
\end{equation}

The matrices $M_k$, $B_k$, and $K_k$ are positive-definite, representing the mass, damping, and stiffness coefficients of the human operator and environment, respectively. The vectors $x_i$, $\dot{x}_i$, and $\ddot{x}_i \in \mathbb{R}^n$ denote the end-effector position, velocity, and acceleration in task space. The nonhomogeneous terms $f_k^*$ are assumed to be bounded.

Substituting the force models and kinematic relationships into the original robot dynamics \eqref{eq:maindyn}, the combined system dynamics become:
\begin{equation}
\label{eq:finaldyn}
\begin{cases} 
    M_m(q_m)\ddot{q}_m + C_m(q_m,\dot{q}_m)\dot{q}_m + G_m(q_m) = \tau_m + \tau_h \\ 
    M_s(q_s)\ddot{q}_s + C_s(q_s,\dot{q}_s)\dot{q}_s + G_s(q_s) = \tau_s + \tau_e 
\end{cases}
\end{equation}
where the combined inertia, Coriolis, and gravity matrices are defined as:
\begin{equation}
\begin{cases}
    M_i(q_i) = M_{qi}(q_i) + J_i^T M_k J_i \\
    C_i(q_i,\dot{q}_i) = C_{qi}(q_i,\dot{q}_i) + J_i^T M_k \dot{J}_i + J_i^T B_k J_i \\
    G_i(q_i) = G_{qi}(q_i) + J_i^T K_k H_i(q_i) 
\end{cases}
\end{equation}
The equivalent joint-space torque contributions from the human and environment are given by:
\begin{equation}
\tau_h = J_m^T f_h^*, \qquad \tau_e = -J_s^T f_e^*.
\end{equation}

With the addition of external forces and joint friction, the dynamic equation takes the following form:
\begin{equation}
\label{eq:finaldyn_f}
M_i(q_i)\ddot{q}_i + C_i(q_i, \dot{q}_i)\dot{q}_i + G_i(q_i) + f_i(q_i, \dot{q}_i) + f_{ci}(\dot{q}_i) = \tau_i + \tau_k
\end{equation}

\section{Controller Design}
\label{sec:control_design}

The teleoperation system is organized around a position-error control
structure, in which the position measurements exchanged between the two
sites are transmitted with constant delays, namely $T_m$ in the forward
(master-to-slave) communication channel and $T_s$ in the reverse
(slave-to-master) channel. Using the delayed positions, the position
tracking errors at the master and slave sides are defined as
\begin{equation}
\label{eq:err-m}
e_m(t)=q_s(t-T_s)-q_m(t),
\end{equation}
\begin{equation}
\label{eq:err-s}
e_s(t)=q_m(t-T_m)-q_s(t),
\end{equation}
where $q_m$ and $q_s$ denote the master and slave joint positions,
respectively.

The control objective is to keep the torques applied to both manipulators
bounded while driving the position tracking error of the master--slave
system to zero, thereby ensuring stability of the closed-loop system.
Toward this end, for $i\in\{m,s\}$, the sliding-mode functions are chosen as
\begin{equation}
\label{eq:slide}
r_i=\dot e_i+\Lambda_i e_i,
\end{equation}
where $\Lambda_i=\Lambda_i^{T}>0$ is a constant positive diagonal matrix.

\begin{figure}[!t]
    \centering
    \includegraphics[width=\linewidth]{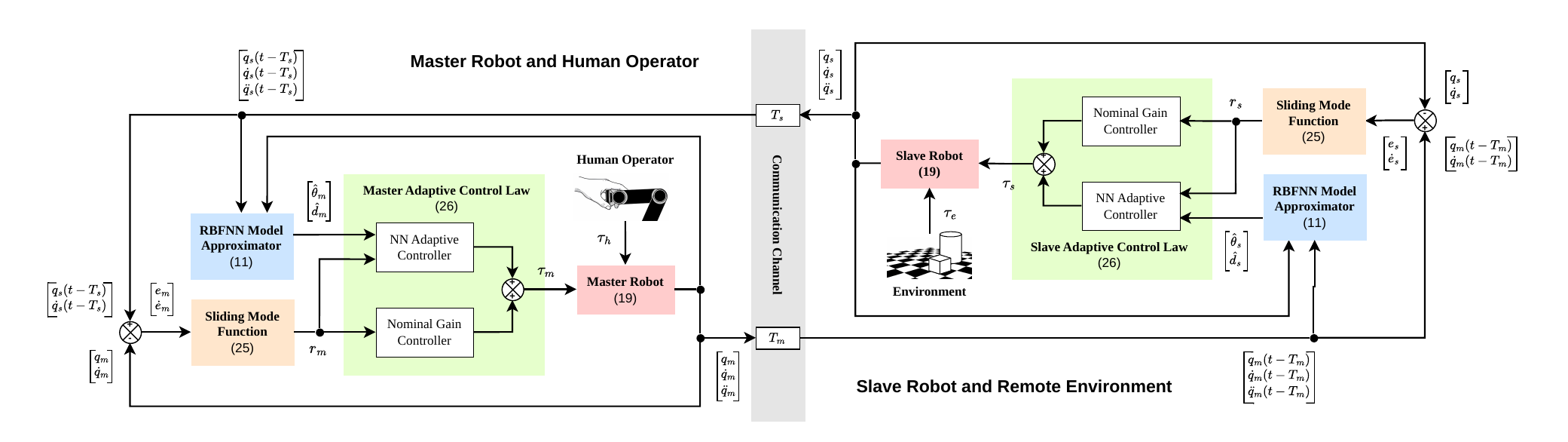}
    \caption{Block diagram of the teleoperation system with the proposed adaptive neural-network controller.}
    \label{fig:blockdiagram}
\end{figure}

An adaptive neural-network controller is designed for each side of the
teleoperation system so that the sliding variables $r_i$ and the estimation
errors remain bounded. The overall control architecture is shown in Fig.~\ref{fig:blockdiagram}. For $i\in\{m,s\}$, the control torque is chosen as
\begin{equation}
\label{eq:control}
\tau_i(t)=k_i\,r_i(t)
+\frac{r_i(t)}{2a_i^{2}}\,\hat\theta_i(t)\,\sigma_i(X_i)^{T}\sigma_i(X_i)
+\frac{r_i(t)}{\norm{r_i(t)}+e^{-a_i t}}\,\hat d_i(t),
\end{equation}
where $k_i>0$ is the feedback gain, $a_i>0$ is a design constant, and
$\hat\theta_i(t)$ and $\hat d_i(t)$ are the on-line estimates of the
neural-network weight norm $\theta_i=\norm{W_i}_F^{2}$ and of the uncertainty
bound $d_i$, respectively.

The first term is a proportional feedback on the sliding variable $r_i$. The second term
compensates the neural-network-approximated model uncertainty; adapting only the
scalar $\hat\theta_i$ rather than the full weight matrix $W_i$ keeps the number
of on-line parameters minimal. The third term is a robust term that rejects the
lumped friction and disturbance bounded by $d_i$.

The estimates are generated by the adaptive update laws
\begin{align}
\label{eq:adapt-theta}
\dot{\hat\theta}_i(t)&=\frac{\lambda_i}{2a_i^{2}}\,
r_i(t)^{T}r_i(t)\,\sigma_i(X_i)^{T}\sigma_i(X_i)-\psi_i\,\hat\theta_i(t),\\[4pt]
\label{eq:adapt-d}
\dot{\hat d}_i(t)&=\frac{\gamma_i\,r_i(t)^{T}r_i(t)}{\norm{r_i(t)}+e^{-a_i t}}
-\upsilon_i\,\hat d_i(t),
\end{align}
where $\lambda_i>0$ and $\gamma_i>0$ are adaptation gains, and $\psi_i>0$ and
$\upsilon_i>0$ are leakage coefficients ($\sigma$-modification) that guarantee
boundedness of the estimates.

\section{Stability Analysis}
\label{sec:stability}

Uniform ultimate boundedness (UUB) of the closed-loop teleoperation system is established by means of a Lyapunov--Krasovskii functional.

Throughout the analysis the indices $i$ and $j$ range over the two sides of the
system and always denote counterparts: $i,j\in\{m,s\}$ with $j\neq i$, so that
$(i,j)\in\{(m,s),(s,m)\}$.


Consider the Lyapunov--Krasovskii functional
\begin{equation}
\label{eq:V}
V(t)=\sum_{i\in\{m,s\}}
\left[\,
\frac12\,r_i^{T}M_i(q_i)\,r_i
+\frac{1}{2\lambda_i}\,\tilde\theta_i^{2}
+\frac{1}{2\gamma_i}\,\tilde d_i^{2}
+V_{Qi}(t)+V_{Ri}(t)
\,\right],
\end{equation}
where the delay-dependent terms are, for each side $i$ and its counterpart $j$,
\begin{align}
\label{eq:VQ}
V_{Qi}(t)&=\int_{t-T_j}^{t}\dot q_j(\rho)^{T}Q_i\,\dot q_j(\rho)\,d\rho,\\[4pt]
\label{eq:VR}
V_{Ri}(t)&=\int_{-T_j}^{0}\!\int_{t+\beta}^{t}
\dot q_j(\rho)^{T}R_i\,\dot q_j(\rho)\,d\rho\,d\beta,
\end{align}
and $Q_i=Q_i^{T}\succ0$, $R_i=R_i^{T}\succ0$ are weighting matrices to be
determined.

\begin{equation}
\label{eta}
\eta_i(t)
=\int_{t-T_j}^{t}\dot{q}_j(\rho)\,d\rho
= q_j(t)-q_j(t-T_j),
\end{equation}


Differentiating the functional \eqref{eq:V} along the closed-loop trajectories
and using the symmetry of $M_i(q_i)$, so that
$\dot r_i^{T}M_i r_i=r_i^{T}M_i\dot r_i$, gives
\begin{equation}
\label{eq:Vdot-raw}
\dot V(t)=\sum_{i\in\{m,s\}}
\left(
\tfrac12\,r_i^{T}\dot M_i\,r_i
+r_i^{T}M_i\,\dot r_i
+\tfrac{1}{\lambda_i}\,\tilde\theta_i\,\dot{\tilde\theta}_i
+\tfrac{1}{\gamma_i}\,\tilde d_i\,\dot{\tilde d}_i
\right)
+\sum_{i\in\{m,s\}}\!\big(\dot V_{Qi}+\dot V_{Ri}\big).
\end{equation}

Substituting the closed-loop dynamics of the sliding variable,
\begin{equation}
\label{eq:closed-loop}
M_i(q_i)\,\dot r_i=\tilde f_i(X_i)-\tau_i+f_i+f_{ci}-C_i(q_i,\dot q_i)\,r_i,
\qquad i\in\{m,s\},
\end{equation}
and grouping the inertia and Coriolis contributions, each kinetic term reduces,
by the  Property 2, to
\begin{equation}
\label{eq:passivity-use}
\tfrac12\,r_i^{T}\dot M_i\,r_i+r_i^{T}M_i\dot r_i
=r_i^{T}\!\big(\tilde f_i(X_i)-\tau_i+f_i+f_{ci}\big)
+\tfrac12\,r_i^{T}\!\big(\dot M_i-2C_i\big)r_i
=r_i^{T}\!\big(\tilde f_i(X_i)-\tau_i+f_i+f_{ci}\big)-r_i^{T}B_i\,r_i .
\end{equation}
Together with $\dot{\tilde\theta}_i=-\dot{\hat\theta}_i$ and
$\dot{\tilde d}_i=-\dot{\hat d}_i$, the derivative \eqref{eq:Vdot-raw} becomes
\begin{equation}
\label{eq:Vdot-sub}
\dot V(t)=\sum_{i\in\{m,s\}}
\left[
r_i^{T}\!\big(\tilde f_i(X_i)-\tau_i+f_i+f_{ci}\big)
-r_i^{T}B_i\,r_i
-\tfrac{1}{\lambda_i}\,\tilde\theta_i\,\dot{\hat\theta}_i
-\tfrac{1}{\gamma_i}\,\tilde d_i\,\dot{\hat d}_i
\right]
+\sum_{i\in\{m,s\}}\!\big(\dot V_{Qi}+\dot V_{Ri}\big).
\end{equation}

Since the delays are constant (Assumption~\ref{ass:delay}), the Leibniz rule applied to \eqref{eq:VQ} and
\eqref{eq:VR} yields, for each side $i$ with counterpart $j$,
\begin{align}
\label{eq:VQdot}
\dot V_{Qi}&=\dot q_j(t)^{T}Q_i\,\dot q_j(t)
-\dot q_j(t-T_j)^{T}Q_i\,\dot q_j(t-T_j),\\[4pt]
\label{eq:VRdot}
\dot V_{Ri}&=T_j\,\dot q_j(t)^{T}R_i\,\dot q_j(t)
-\int_{t-T_j}^{t}\dot q_j(\rho)^{T}R_i\,\dot q_j(\rho)\,d\rho .
\end{align}

Writing $\sigma_i\equiv\sigma_i(X_i)$ for brevity and inserting the
neural-network decomposition \eqref{eq:rbf}, $\tilde f_i=W_i^{T}\sigma_i+\delta_i$,
into \eqref{eq:Vdot-sub} gives
\begin{equation}
\label{eq:Vdot-nn}
\dot V(t)=\sum_{i\in\{m,s\}}
\left[
r_i^{T}\!\big(W_i^{T}\sigma_i+\delta_i-\tau_i+f_i+f_{ci}\big)
-r_i^{T}B_i r_i
-\tfrac{1}{\lambda_i}\tilde\theta_i\dot{\hat\theta}_i
-\tfrac{1}{\gamma_i}\tilde d_i\dot{\hat d}_i
\right]
+\sum_{i\in\{m,s\}}\!\big(\dot V_{Qi}+\dot V_{Ri}\big).
\end{equation}

By the Cauchy--Schwarz and Young inequalities (Lemma~\ref{lem:young}), with
$\theta_i=\norm{W_i}_F^{2}$,
\begin{align}
\label{eq:nn-bound}
r_i^{T}W_i^{T}\sigma_i
&\le\norm{r_i}\,\norm{W_i}_F\,\norm{\sigma_i}
\le\frac{1}{2a_i^{2}}\,r_i^{T}r_i\,\theta_i\,\sigma_i^{T}\sigma_i+\frac12 a_i^{2},\\[4pt]
\label{eq:dist-bound}
r_i^{T}\!\big(\delta_i+f_{ci}+f_i\big)
&\le\norm{r_i}\,d_i,
\end{align}
where \eqref{eq:dist-bound} follows from Assumption~\ref{ass:disturbance}.
Substituting \eqref{eq:nn-bound}--\eqref{eq:dist-bound} into
\eqref{eq:Vdot-nn} yields
\begin{equation}
\label{eq:Vdot-40}
\dot V(t)\le\sum_{i\in\{m,s\}}
\left[
\frac{1}{2a_i^{2}}\,r_i^{T}r_i\,\theta_i\,\sigma_i^{T}\sigma_i
-r_i^{T}\tau_i
+\norm{r_i}\,d_i+\tfrac12 a_i^{2}
-r_i^{T}B_i r_i
-\tfrac{1}{\lambda_i}\tilde\theta_i\dot{\hat\theta}_i
-\tfrac{1}{\gamma_i}\tilde d_i\dot{\hat d}_i
\right]
+\sum_{i\in\{m,s\}}\!\big(\dot V_{Qi}+\dot V_{Ri}\big).
\end{equation}

Left-multiplying the control law \eqref{eq:control} by $-r_i^{T}$ and
substituting into \eqref{eq:Vdot-40}, the neural-network terms combine through
$\tilde\theta_i=\theta_i-\hat\theta_i$ and the robust term splits through
$\hat d_i=d_i-\tilde d_i$, which gives
\begin{equation}
\label{eq:Vdot-43}
\begin{aligned}
\dot V(t)\le{}&\sum_{i\in\{m,s\}}\Bigg[
-k_i\,r_i^{T}r_i+\tfrac12 a_i^{2}-r_i^{T}B_i r_i
+\frac{1}{\lambda_i}\tilde\theta_i\!\left(\frac{\lambda_i}{2a_i^{2}}r_i^{T}r_i\,\sigma_i^{T}\sigma_i-\dot{\hat\theta}_i\right)\\[2pt]
&\hphantom{\sum_{i\in\{m,s\}}\Bigg[}
+\frac{1}{\gamma_i}\tilde d_i\!\left(\frac{\gamma_i\,r_i^{T}r_i}{\norm{r_i}+e^{-a_i t}}-\dot{\hat d}_i\right)
+\norm{r_i}\,d_i-\frac{r_i^{T}r_i}{\norm{r_i}+e^{-a_i t}}\,d_i
\Bigg]
+\sum_{i\in\{m,s\}}\!\big(\dot V_{Qi}+\dot V_{Ri}\big).
\end{aligned}
\end{equation}

The adaptive laws \eqref{eq:adapt-theta}--\eqref{eq:adapt-d} are selected such that the two parenthesized terms in \eqref{eq:Vdot-43} reduce to the leakage contributions \(-\psi_i\hat\theta_i\) and \(-\upsilon_i\hat d_i\), which yield the $\sigma$-modification terms in the Lyapunov derivative.

Bounding the residual integral in \eqref{eq:VRdot} by Jensen's inequality
(Lemma~\ref{lem:jensen}),
\begin{equation}
\label{eq:jensen-use}
\int_{t-T_j}^{t}\dot q_j(\rho)^{T}R_i\,\dot q_j(\rho)\,d\rho
\;\ge\;\frac{1}{T_j}\,\eta_i^{T}R_i\,\eta_i ,
\end{equation}
and recalling the delay-loaded weight $S_i=Q_i+T_jR_i$, the single- and double-integral terms of side $i$ combine into
\begin{equation}
\label{eq:Kraso-pair}
\dot V_{Qi}+\dot V_{Ri}
\;\le\;
\dot q_j(t)^{T}S_i\,\dot q_j(t)
-\dot q_j(t-T_j)^{T}Q_i\,\dot q_j(t-T_j)
-\frac{1}{T_j}\,\eta_i^{T}R_i\,\eta_i .
\end{equation}

Substituting the leakage identities \eqref{eq:adapt-theta},\eqref{eq:adapt-d} and \eqref{eq:Kraso-pair}  into \eqref{eq:Vdot-43} yields
\begin{equation}
\label{eq:Vdot-46}
\begin{aligned}
\dot V(t)\le{}&\sum_{i\in\{m,s\}}
\left[
-k_i\,r_i^{T}r_i-r_i^{T}B_i r_i
+\frac{\psi_i}{\lambda_i}\tilde\theta_i\hat\theta_i
+\frac{\upsilon_i}{\gamma_i}\tilde d_i\hat d_i
+\norm{r_i}\,d_i-\frac{r_i^{T}r_i}{\norm{r_i}+e^{-a_i t}}\,d_i
+\tfrac12 a_i^{2}
\right]\\
&+\sum_{i\in\{m,s\}}
\left[
\dot q_j(t)^{T}S_i\,\dot q_j(t)
-\dot q_j(t-T_j)^{T}Q_i\,\dot q_j(t-T_j)
-\frac{1}{T_j}\,\eta_i^{T}R_i\,\eta_i
\right].
\end{aligned}
\end{equation}

Writing $\hat\theta_i=\theta_i-\tilde\theta_i$ and applying Young's inequality
(Lemma~\ref{lem:young}) in the form $\tilde\theta_i\theta_i\le\tfrac12\theta_i^2+\tfrac12\tilde\theta_i^2$,
\begin{equation}
\label{eq:sq-theta}
\frac{\psi_i}{\lambda_i}\,\tilde\theta_i\hat\theta_i
=\frac{\psi_i}{\lambda_i}\,\tilde\theta_i\big(\theta_i-\tilde\theta_i\big)
\;\le\;\frac{\psi_i}{2\lambda_i}\,\theta_i^{2}-\frac{\psi_i}{2\lambda_i}\,\tilde\theta_i^{2},
\end{equation}
and, identically, with $\hat d_i=d_i-\tilde d_i$,
\begin{equation}
\label{eq:sq-d}
\frac{\upsilon_i}{\gamma_i}\,\tilde d_i\hat d_i
\;\le\;\frac{\upsilon_i}{2\gamma_i}\,d_i^{2}-\frac{\upsilon_i}{2\gamma_i}\,\tilde d_i^{2}.
\end{equation}
The robust residual is collapsed by combining its two terms over a common
denominator,
\begin{equation}
\label{eq:dist-collapse}
\norm{r_i}\,d_i-\frac{r_i^{T}r_i}{\norm{r_i}+e^{-a_i t}}\,d_i
=\frac{\norm{r_i}\,e^{-a_i t}}{\norm{r_i}+e^{-a_i t}}\,d_i
\;\le\;e^{-a_i t}\,d_i,
\end{equation}
where the last step uses $\frac{\norm{r_i}}{\norm{r_i}+e^{-a_i t}}\le1$.

Substituting \eqref{eq:sq-theta}--\eqref{eq:dist-collapse} into
\eqref{eq:Vdot-46} and grouping terms gives
\begin{equation}
\label{eq:Vdot-49}
\begin{aligned}
\dot V(t)\le{}&
-\sum_{i\in\{m,s\}}\left(
k_i\,r_i^{T}r_i+r_i^{T}B_i r_i
+\frac{\psi_i}{2\lambda_i}\tilde\theta_i^{2}
+\frac{\upsilon_i}{2\gamma_i}\tilde d_i^{2}
\right)\\[2pt]
&+\sum_{i\in\{m,s\}}\left(
\dot q_j(t)^{T}S_i\,\dot q_j(t)
-\dot q_j(t-T_j)^{T}Q_i\,\dot q_j(t-T_j)
-\frac{1}{T_j}\,\eta_i^{T}R_i\,\eta_i
\right)
+b_0(t),
\end{aligned}
\end{equation}
where the residual collects the constant and exponentially decaying terms,
\begin{equation}
\label{eq:b0}
b_0(t):=\sum_{i\in\{m,s\}}\left(
\frac{\psi_i}{2\lambda_i}\theta_i^{2}
+\frac{\upsilon_i}{2\gamma_i}d_i^{2}
+e^{-a_i t}d_i
+\tfrac12 a_i^{2}
\right)\le\bar b_0 ,
\end{equation}
and $\bar b_0=\sum_{i\in\{m,s\}}\big(\tfrac{\psi_i}{2\lambda_i}\theta_i^{2}+\tfrac{\upsilon_i}{2\gamma_i}d_i^{2}+d_i+\tfrac12 a_i^{2}\big)$.


With $\He(X):=X+X^{T}$ and the abbreviations $v_m=\dot q_m(t)$,
$w=\dot q_s(t-T_s)$, $v_s=\dot q_s(t)$, $u=\dot q_m(t-T_m)$, the sliding-surface
definitions $r_i=\dot e_i+\Lambda_i e_i$ give, on the trajectories,
\begin{equation}
\label{eq:residuals}
\Pi_m:=w-v_m+\Lambda_m e_m-r_m\equiv0,
\qquad
\Pi_s:=u-v_s+\Lambda_s e_s-r_s\equiv0 .
\end{equation}
Since $\Pi_m\equiv0$ and $\Pi_s\equiv0$, for any free-weighting matrices
$N_1,N_2,N_3$ and $L_1,L_2,L_3\in\R^{n\times n}$ the quantities
\begin{equation}
\label{eq:injection}
J_m=2\big(v_m^{T}N_1^{T}+r_m^{T}N_2^{T}+e_m^{T}N_3^{T}\big)\Pi_m=0,
\qquad
J_s=2\big(v_s^{T}L_1^{T}+r_s^{T}L_2^{T}+e_s^{T}L_3^{T}\big)\Pi_s=0.
\end{equation}

Collect the variables appearing in each residual into
\begin{equation}
\label{eq:chi}
\chi_m=\begin{bmatrix}r_m\\ v_m\\ w\\ e_m\end{bmatrix},
\qquad
\chi_s=\begin{bmatrix}r_s\\ v_s\\ u\\ e_s\end{bmatrix}.
\end{equation}

Adding $J_m+J_s$ to \eqref{eq:Vdot-49} and collecting the quadratic and bilinear
terms into $\chi_m^{T}\Theta_m\chi_m+\chi_s^{T}\Theta_s\chi_s$ yields
\begin{equation}
\label{eq:Theta-m}
\Theta_m=
\begin{bmatrix}
-(k_mI+B_h)-\He(N_2) & -N_1-N_2^{T} & N_2^{T} & N_2^{T}\Lambda_m-N_3\\[2pt]
\star & S_s-\He(N_1) & N_1^{T} & N_1^{T}\Lambda_m-N_3\\[2pt]
\star & \star & -Q_m & N_3\\[2pt]
\star & \star & \star & \He(N_3^{T}\Lambda_m)
\end{bmatrix},
\quad S_s=Q_s+T_mR_s,
\end{equation}
\begin{equation}
\label{eq:Theta-s}
\Theta_s=
\begin{bmatrix}
-(k_sI+B_e)-\He(L_2) & -L_1-L_2^{T} & L_2^{T} & L_2^{T}\Lambda_s-L_3\\[2pt]
\star & S_m-\He(L_1) & L_1^{T} & L_1^{T}\Lambda_s-L_3\\[2pt]
\star & \star & -Q_s & L_3\\[2pt]
\star & \star & \star & \He(L_3^{T}\Lambda_s)
\end{bmatrix},
\quad S_m=Q_m+T_sR_m,
\end{equation}
where $\star$ denotes the symmetric transpose of the corresponding upper block.
The $(2,2)$ block $S_s-\He(N_1)$ shows that the destabilizing weight is
absorbed by the free matrix $N_1$ rather than by $Q_m$. Therefore, selecting
$\He(N_1)\succ S_s$ directly renders this block negative. The $(4,4)$ block
$\He(N_3^{T}\Lambda_m)$ embeds the sliding-surface relation; choosing
$N_3=-\mu\Lambda_m^{-1}$ with $\mu>0$ gives
$\He(N_3^{T}\Lambda_m)=-2\mu I\prec0$. The drift variables
$\eta_m,\eta_s$ do not appear in the residual terms and remain uncoupled,
contributing the negative blocks
$-\tfrac{1}{T_s}R_m$ and $-\tfrac{1}{T_m}R_s$.

With the injection in place, \eqref{eq:Vdot-49} becomes
\begin{equation}
\label{eq:Vdot-final}
\dot V(t)\le
\chi_m^{T}\Theta_m\chi_m+\chi_s^{T}\Theta_s\chi_s
-\frac{1}{T_s}\eta_m^{T}R_m\eta_m-\frac{1}{T_m}\eta_s^{T}R_s\eta_s
-\sum_{i\in\{m,s\}}\!\left(\frac{\psi_i}{2\lambda_i}\tilde\theta_i^{2}
+\frac{\upsilon_i}{2\gamma_i}\tilde d_i^{2}\right)+b_0(t),
\end{equation}

\label{thm:uub}
 If, for the delays $T_m,T_s$, there
exist symmetric matrices $Q_m,Q_s,R_m,R_s\succ0$ and free-weighting matrices
$N_1,N_2,N_3,L_1,L_2,L_3$ such that
\begin{equation}
\label{eq:lmi}
\Theta_m(T_m,T_s)\prec0
\qquad\text{and}\qquad
\Theta_s(T_m,T_s)\prec0,
\end{equation}
with $\Theta_m,\Theta_s$ given by \eqref{eq:Theta-m}--\eqref{eq:Theta-s}, then
all closed-loop signals $r_i,e_i,\tilde\theta_i,\tilde d_i$ ($i\in\{m,s\}$) are
uniformly ultimately bounded in the sense of Definition~\ref{def:uub}.

Collect the closed-loop variables into
\begin{equation}
\label{eq:Z}
Z=\begin{bmatrix}\chi_m^{\top}&\chi_s^{\top}&\eta_m^{\top}&\eta_s^{\top}
&\tilde\theta_m&\tilde\theta_s&\tilde d_m&\tilde d_s\end{bmatrix}^{\top}.
\end{equation}
Under \eqref{eq:lmi} the quadratic forms satisfy
$\chi_m^{\top}\Theta_m\chi_m\le-\lambda_{\min}(-\Theta_m)\lVert\chi_m\rVert^2$
and
$\chi_s^{\top}\Theta_s\chi_s\le-\lambda_{\min}(-\Theta_s)\lVert\chi_s\rVert^2$,
while $R_m,R_s\succ0$ and $\psi_i,\upsilon_i,\lambda_i,\gamma_i>0$ render the
Jensen and adaptive-error terms in \eqref{eq:Vdot-final} negative. Hence there
is a constant
\begin{equation}
\alpha=\min\Big\{\lambda_{\min}(-\Theta_m),\ \lambda_{\min}(-\Theta_s),\
\tfrac{1}{T_s}\lambda_{\min}(R_m),\ \tfrac{1}{T_m}\lambda_{\min}(R_s),\
\min_{i}\tfrac{\psi_i}{2\lambda_i},\ \min_{i}\tfrac{\upsilon_i}{2\gamma_i}\Big\}>0
\label{eq:alpha}
\end{equation}
such that \eqref{eq:Vdot-final} gives
\begin{equation}
\dot V(t)\ \le\ -\alpha\lVert Z(t)\rVert^2+b_0(t)\ \le\ -\alpha\lVert Z(t)\rVert^2+\bar b_0 .
\label{eq:vdot-Z}
\end{equation}

Collect the \emph{core} pointwise variables into
\begin{equation}
\zeta(t)\;:=\;\big(\,r_m,\ r_s,\ \tilde\theta_m,\ \tilde\theta_s,\ \tilde d_m,\ \tilde d_s\,\big)(t),
\label{eq:zeta}
\end{equation}
where \(x_t(\theta):=x(t+\theta)\), \(\theta\in[-\bar T,0]\), denotes the
history segment of the closed-loop state, and
\[
\lVert x_t\rVert_c:=\sup_{\theta\in[-\bar T,0]}\lVert x(t+\theta)\rVert ,
\]
with \(\bar T:=\max\{T_m,T_s\}\). Since \(V\) is a Lyapunov--Krasovskii functional due to the history-dependent terms \(V_{Q_i}\) and \(V_{R_i}\), the UUB property is
established by deriving suitable lower and upper bounds for \(V\) and
showing decrease outside a compact set.

\medskip
\noindent\textbf{Step 1}
Discarding the nonnegative Krasovskii integrals $V_{Q_i},V_{R_i}\ge 0$ in
\eqref{eq:V} and using Property~1,
\begin{equation}
V(t)\;\ge\;\underline V(t)
:=\sum_{i\in\{m,s\}}\!\Big(\tfrac12\lambda_{\min}\{M_i\}\,\lVert r_i\rVert^2
+\tfrac{1}{2\lambda_i}\tilde\theta_i^{\,2}
+\tfrac{1}{2\gamma_i}\tilde d_i^{\,2}\Big)
\;\ge\; c_1\,\lVert\zeta(t)\rVert^2 =: u_1\!\big(\lVert\zeta(t)\rVert\big),
\label{eq:lower}
\end{equation}
with
\begin{equation}
c_1:=\tfrac12\min_{i\in\{m,s\}}\big\{\lambda_{\min}\{M_i\},\ \lambda_i^{-1},\ \gamma_i^{-1}\big\}>0,
\qquad u_1(s)=c_1 s^2\in\mathcal K_\infty .
\label{eq:c1}
\end{equation}

\medskip
\noindent\textbf{Step 2}
The kinetic and adaptive terms are bounded pointwise, and the Krasovskii
terms are bounded over the delay window directly from their definitions,
\begin{equation}
\tfrac12 r_i^{\top}M_i r_i\le\tfrac12\lambda_{\max}\{M_i\}\lVert r_i\rVert^2,\qquad
V_{Q_i}(t)\le T_j\lambda_{\max}\{Q_i\}\!\!\sup_{\sigma\in[t-T_j,t]}\!\!\lVert\dot q_j(\sigma)\rVert^2,\qquad
V_{R_i}(t)\le\tfrac{T_j^{2}}{2}\lambda_{\max}\{R_i\}\!\!\sup_{\sigma\in[t-T_j,t]}\!\!\lVert\dot q_j(\sigma)\rVert^2 .
\label{eq:kras-ub}
\end{equation}
Since $\lVert x_t\rVert_c$ dominates every component norm and windowed
supremum above, summing over $i\in\{m,s\}$ gives a single
history-dependent upper bound
\begin{equation}
V(t)\;\le\;c_2\,\lVert x_t\rVert_c^{2}=:u_2\!\big(\lVert x_t\rVert_c\big),
\qquad u_2(s)=c_2 s^2\in\mathcal K_\infty,
\label{eq:upper}
\end{equation}
\begin{equation}
c_2:=\!\!\sum_{i\in\{m,s\}}\!{\tfrac12\lambda_{\max}\{M_i\}
+\tfrac{1}{2\lambda_i}+\tfrac{1}{2\gamma_i}}
+{T_j\,\lambda_{\max}\{Q_i\}+\tfrac{T_j^{2}}{2}\lambda_{\max}\{R_i\}}.
\label{eq:c2}
\end{equation}
The history-dependent terms are therefore bounded by the closed-loop
history norm \(\|x_t\|_c\).

\medskip
\noindent\textbf{Step 3}
Since $\zeta$ is a sub-vector of $Z$ in \eqref{eq:Z},
$\lVert Z\rVert^2\ge\lVert\zeta\rVert^2$, so \eqref{eq:vdot-Z} gives a
decrease in the current core state,
\begin{equation}
\dot V(t)\;\le\;-\alpha\lVert Z(t)\rVert^2+\bar b_0
\;\le\;-\alpha\lVert\zeta(t)\rVert^2+\bar b_0
\;\le\;-\tfrac{\alpha}{2}\lVert\zeta(t)\rVert^2
\quad\text{whenever}\quad
\lVert\zeta(t)\rVert\ge\mu:=\sqrt{\tfrac{2\bar b_0}{\alpha}},
\label{eq:decrease}
\end{equation}
with $\alpha>0$ from \eqref{eq:alpha} and $\bar b_0$ from \eqref{eq:b0};
thus $W_3(\zeta):=\tfrac{\alpha}{2}\lVert\zeta\rVert^2$ is continuous
positive-definite and $\dot V<0$ outside the ball
$\lVert\zeta\rVert\le\mu$. Therefore, the Lyapunov--Krasovskii UUB conditions are satisfied:
\begin{equation*}
u_1\!\big(\lVert\zeta(t)\rVert\big)\ \le\ V(t)\ \le\ u_2\!\big(\lVert x_t\rVert_c\big),
\qquad
\dot V(t)\le -W_3\!\big(\zeta(t)\big)\ \ \forall\,\lVert\zeta(t)\rVert\ge\mu,
\qquad u_1,u_2\in\mathcal K_\infty .
\end{equation*}
Hence $V$ is bounded for all $t\ge0$, and $\zeta$ is uniformly ultimately
bounded with
\begin{equation}
\limsup_{t\to\infty}\lVert\zeta(t)\rVert\ \le\ u_1^{-1}\!\big(u_2(\mu)\big)
=\sqrt{\tfrac{c_2}{c_1}}\;\mu
=\sqrt{\tfrac{2\,c_2\,\bar b_0}{c_1\,\alpha}} .
\label{eq:ubound}
\end{equation}
Equivalently,
\begin{equation}
\limsup_{t\to\infty}V(t)\ \le\ \bar V:=u_2(\mu)=c_2\mu^2=\frac{2\,c_2\,\bar b_0}{\alpha}.
\label{eq:Vbar}
\end{equation}

\smallskip

\medskip
\noindent\textbf{Step 4}
By \eqref{eq:lower} and \eqref{eq:Vbar}, \(r_i,\tilde\theta_i,\tilde d_i\)
are uniformly ultimately bounded, with
\begin{equation}
\limsup_{t\to\infty}\lVert r_i\rVert\le\sqrt{\tfrac{2\bar V}{\lambda_{\min}\{M_i\}}},\qquad
\limsup_{t\to\infty}|\tilde\theta_i|\le\sqrt{2\lambda_i\bar V},\qquad
\limsup_{t\to\infty}|\tilde d_i|\le\sqrt{2\gamma_i\bar V},\qquad i\in\{m,s\},
\label{eq:comp}
\end{equation}
each proportional to \(\sqrt{\bar b_0/\alpha}\) through
\(\bar V=2c_2\bar b_0/\alpha\). Since \(\Lambda_i\succ0\), the error
dynamics are exponentially stable with
respect to the input \(r_i\). Hence, the UUB property of \(r_i\) implies
the UUB property of \(e_i\) and \(\dot e_i\), where
\begin{equation}
\limsup_{t\to\infty}\lVert e_i\rVert\le
\frac{1}{\lambda_{\min}\{\Lambda_i\}}
\limsup_{t\to\infty}\lVert r_i\rVert
\le
\frac{1}{\lambda_{\min}\{\Lambda_i\}}
\sqrt{\tfrac{2\bar V}{\lambda_{\min}\{M_i\}}}.
\label{eq:ebound}
\end{equation}
Therefore, all closed-loop signals \(r_i,e_i,\tilde\theta_i,\tilde d_i\)
(\(i\in\{m,s\}\)) are uniformly ultimately bounded according to
Definition~\ref{def:uub}. 
\section{Simulation Results}
\label{sec:results}


The feasibility of the proposed delay-dependent stability condition is evaluated for different communication delays by solving the associated LMIs. The negative definiteness of $\Theta$ is verified using the value of $-\lambda_{\max}(\Theta)$, where positive values indicate a feasible LMI solution. As shown in Table~\ref{tab:LMI_feasibility}, the proposed stability condition remains feasible up to approximately $T=0.25\,\mathrm{s}$, while $T=0.256\,\mathrm{s}$ is close to the feasibility boundary. For larger delays, such as $T=0.30\,\mathrm{s}$, the LMI condition becomes infeasible. In the following simulations, a communication delay of $T=0.1\,\mathrm{s}$ is considered.

\begin{table}[H]
\centering
\caption{Feasibility analysis of the delay-dependent LMI condition for different communication delays.}
\label{tab:LMI_feasibility}
\begin{tabular}{ccc}
\hline
$T$ (s) & $-\lambda_{\max}(\Theta)$ & LMI status \\
\hline
0.05  & 442.1   & Feasible \\
0.10  & 330.6   & Feasible \\
0.15  & 209.2   & Feasible \\
0.20  & 107.1   & Feasible \\
0.256 & $\approx 0$ & Boundary of feasibility \\
0.30  & -0.26   & Infeasible \\
\hline
\end{tabular}
\end{table}


The proposed controller is verified in MATLAB/Simulink. The master and the slave are each modeled as a 2-DOF, revolute-joint manipulator. The two arms are coupled only through a delayed communication channel. The scenario below combines a bounded communication delay, joint friction, an external disturbance, and an operator force pushing on the master. The system is controlled by a neural network adaptive controller; the last part of the section then looks at how the controller gains shape the response.

Neglecting the moment of inertia of the rod, the mathematical model of the 2-link robot based on (\ref{eq:finaldyn}) is as follows:
\begin{equation}
M_{q_i}(q_i) = 
\begin{bmatrix}
m_{i1}l_{i1}^2 + m_{i2}l_{i1}^2 + m_{i2}l_{i2}^2 + 2m_{i2}l_{i1}l_{i2} \cos(q_{i2}) & m_{i2}l_{i2}^2 + m_{i2}l_{i1}l_{i2} \cos(q_{i2}) \\
m_{i2}l_{i2}^2 + m_{i2}l_{i1}l_{i2} \cos(q_{i2}) & m_{i2}l_{i2}^2
\label{eq:M_qi_Results}
\end{bmatrix}
\end{equation}

\begin{equation}
C_{q_i}(q_i, \dot{q}_i) = 
\begin{bmatrix}
-m_{i2}l_{i1}l_{i2} \dot{q}_{i2} \cos(q_{i2}) & -m_{i2}l_{i1}l_{i2} (\dot{q}_{i1} + \dot{q}_{i2}) \sin(q_{i2}) \\
m_{i2}l_{i1}l_{i2} \dot{q}_{i1} \sin(q_{i2}) & 0
\end{bmatrix}
\end{equation}

\begin{equation}
G_{q_i}(q_i) = 
\begin{bmatrix}
(m_{i1}l_{i2} + m_{i2}l_{i1})g \cos(q_{i1}) + m_{i2}l_{i2}g \cos(q_{i1} + q_{i2}) \\
m_{i2}l_{i2}g \cos(q_{i1} + q_{iu})
\end{bmatrix}
\end{equation}

\begin{equation}
H_i = 
\begin{bmatrix}
l_{i1} \cos(q_{i1}) + l_{i2} \cos(q_{i1} + q_{i2}) & l_{i1} \sin(q_{i1}) + l_{i2} \sin(q_{i1} + q_{i2})
\end{bmatrix}
\end{equation}

\begin{equation}
J_i(q_i) = 
\begin{bmatrix}
-l_{i1} \sin(q_{i1}) - l_{i2} \sin(q_{i1} + q_{i2}) & -l_{i2} \sin(q_{i1} + q_{i2}) \\
-l_{i2} \sin(q_{i1} + q_{i2}) & l_{i2} \cos(q_{i1} + q_{i2})
\end{bmatrix}.
\end{equation}

In addition, in the simulation, the external disturbances acting on the master and slave robots are set as 
\begin{equation}
f_i(\dot{q}_i, \dot{q}_i) = \begin{bmatrix} 0.1q_{i1}\dot{q}_{i1}\sin(t) & 0.1q_{i2}\dot{q}_{i2}\sin(t) ^T.
\end{bmatrix}
\end{equation}

The internal friction of the master robot is defined as 
\begin{equation}
f_{cm}(\dot{q}_m) = \bigl[f_{d1}\dot{q}_{m1} + k_1 \sin(q_{m1}),\quad f_{d2}\dot{q}_{m2} + k_2 \sin(q_{m2})\bigr]^T,
\end{equation}

where \(f_{d1}, f_{d2}, k_1, k_2\) are constants. Similarly, the internal friction of the slave robot is given by the following equation with \(f_{d3}, f_{d4}, k_3, k_4\) being constant parameters.

\begin{equation}
f_{cs}(\dot{q}_s) = \bigl[f_{d3}\dot{q}_{s1} + k_3 \sin(q_{s1}),\quad f_{d4}\dot{q}_{s2} + k_4 \sin(q_{s2})\bigr]^T,
\end{equation}

Parameters of the simulation are set according to Tables \ref{tab:parameters} and \ref{tab:params2}:

\begin{table}[H]
\centering
\caption{Master-slave robot, operator, environment, and communication parameters}
\label{tab:parameters}
\begin{tabular}{c|c|c|c}
\hline
\textbf{Parameter} & \textbf{Value} & \textbf{Parameter} & \textbf{Value} \\
\hline
$m_{m1}$ & 1.5 kg & $m_{s1}$ & 1.5 kg \\
$l_{m1}$ & 0.6 m & $l_{s1}$ & 0.6 m \\
$m_{m2}$ & 1.0 kg & $m_{s2}$ & 1.0 kg \\
$l_{m2}$ & 0.4 m & $l_{s2}$ & 0.4 m \\
\hline
$M_h$ & 0.1 kg & $M_e$ & 0.1 kg \\
$B_h$ & 10 Ns/m & $B_e$ & 10 Ns/m \\
$K_h$ & 1000 N/m & $K_e$ & 1000 N/m \\
\hline
$q_{m1}(0)$ & $0.4\pi$ rad & $q_{s1}(0)$ & $0.25\pi$ rad \\
$q_{m2}(0)$ & $0.2\pi$ rad & $q_{s2}(0)$ & $0.1\pi$ rad \\
$T_m$ & 0.1 s & $T_s$ & 0.1 s \\
\hline
\end{tabular}
\end{table}

\begin{table}[htbp]
\centering
\caption{Friction and disturbance parameters}
\label{tab:params2}
\begin{tabular}{c|c}
\hline
\textbf{Parameter} & \textbf{Value} \\
\hline
Viscous friction coefficients $[f_{d1}, \ldots, f_{d4}]$ & $1, 2, 3, 3$ \\
Coulomb friction coefficients $[k_1, \ldots, k_4]$ & $3, 2, 4, 6$ \\
\hline
\end{tabular}
\end{table}

\begin{figure}[H]
    \centering
    \includegraphics[width=0.5\linewidth]{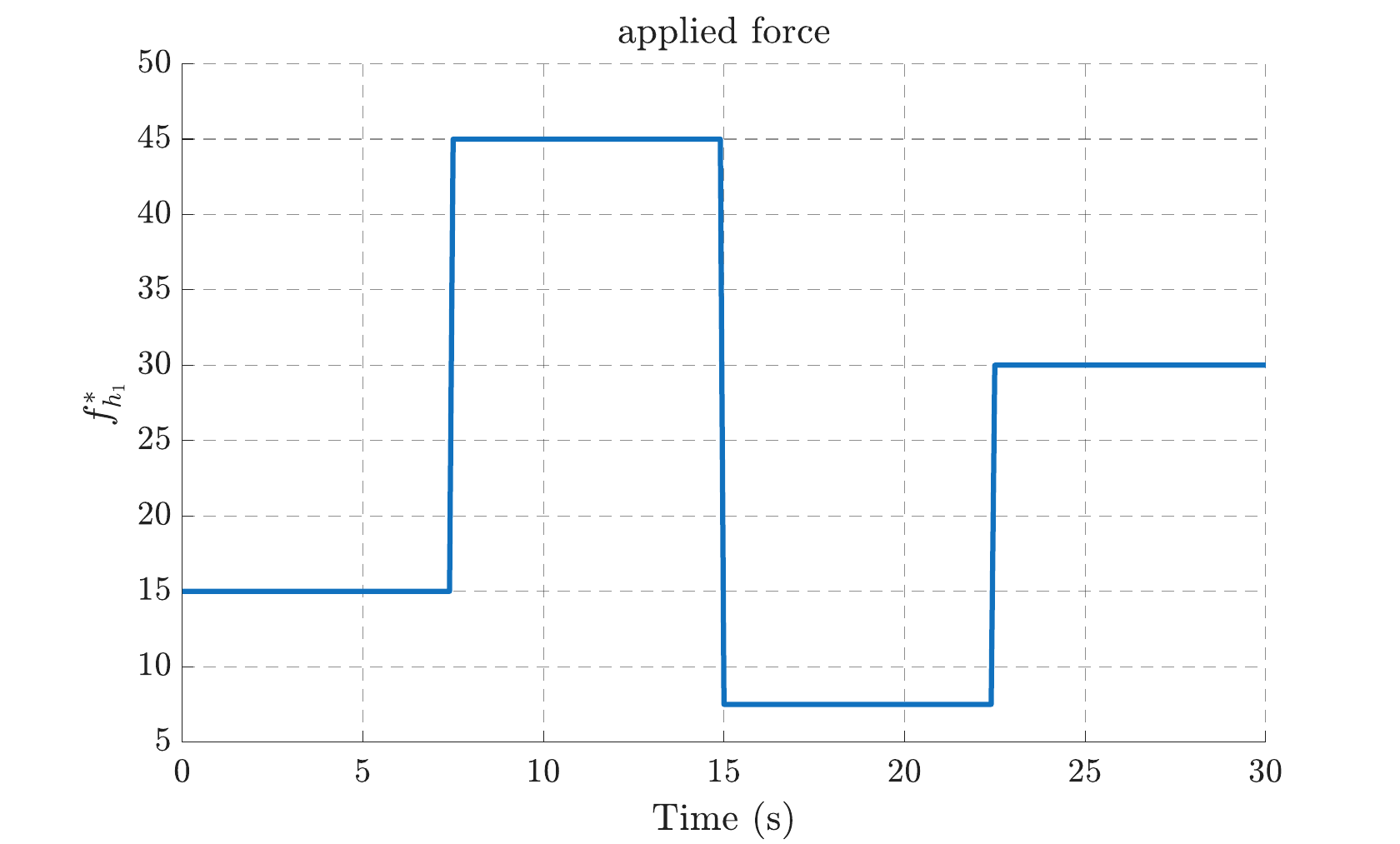}
    \caption{Applied human force}
    \label{fig:human-force}
\end{figure}

An operator force (in one dimension) is applied to the master robot as per Fig (\ref{fig:human-force}), which is applied to the system as \(-J^T f_{h_1}^*\) (the whole force is applied as $\begin{bmatrix}
    f_{h_1}^*\\0
\end{bmatrix}$.

The master and slave controllers are given by Equation \ref{eq:control} with adaptive laws \ref{eq:adapt-theta} and \ref{eq:adapt-d}. After tuning, the controller parameters are selected as
$k_m = k_s = \operatorname{diag}(100, 100), \quad a_m = a_s = 5,$
and the sliding mode parameters are
$\Lambda_m = \Lambda_s = 5I$
The adaptive law parameters are
$
\gamma_m = \gamma_s = \lambda_m = \lambda_s = 1, \quad
\psi_m = \psi_s = 0.5, \quad \nu_m = \nu_s = 0.5.
$
The neural network contains 50 nodes with centers uniformly distributed over \([-2, 2]\).

\begin{figure}[H]
    \centering
    \includegraphics[width=0.8\linewidth]{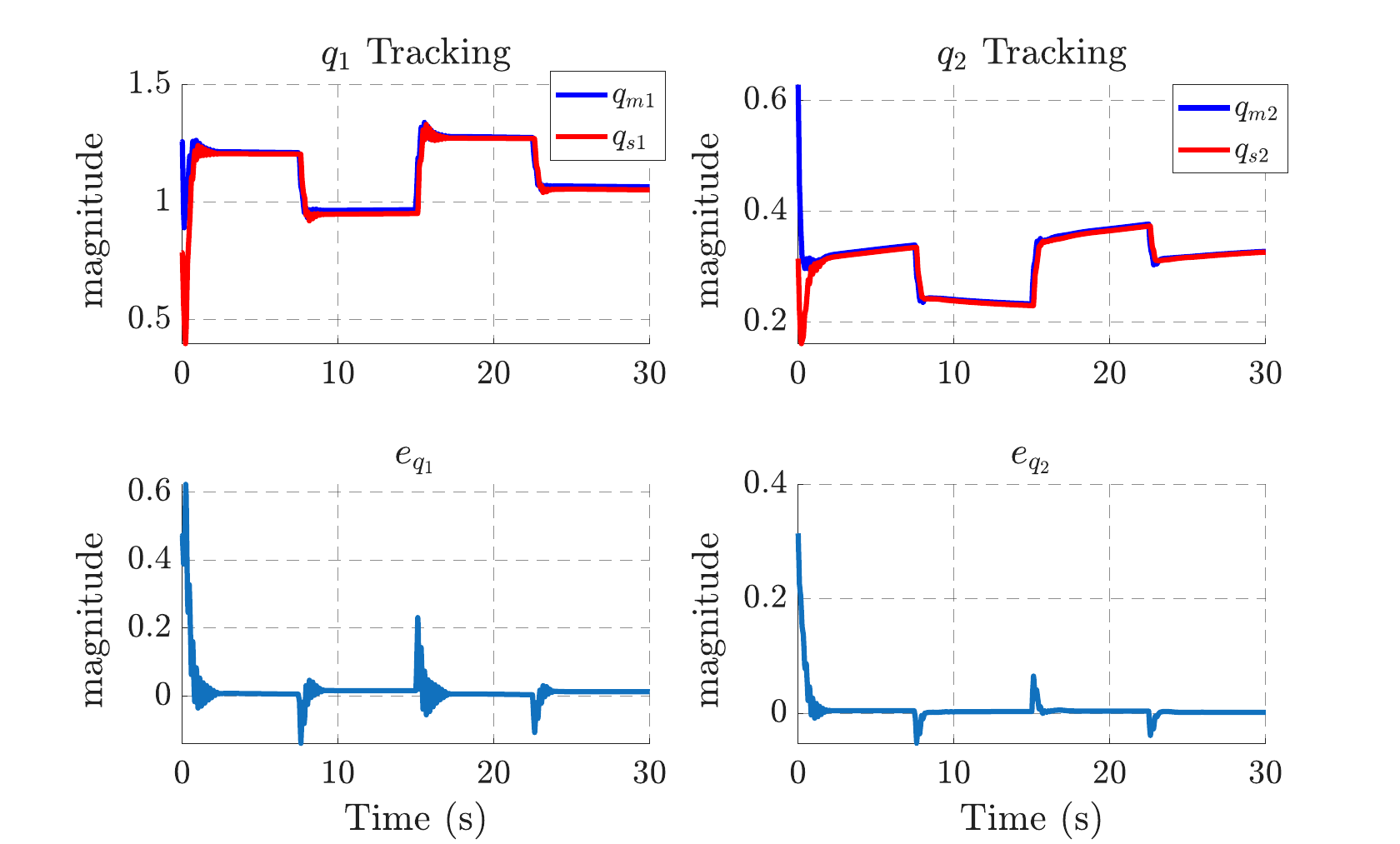}
    \caption{Master-slave joint tracking and error performance}
    \label{fig:tracking}
\end{figure}

The tracking performance of the designed controller can be seen in Figure (\ref{fig:tracking})

The top row shows the master and slave joint angles, and the bottom row shows the synchronization error \( e_j = q_{mj} - q_{sj} \). it can be seen that the slave catches up to the master within about \( 2 \)–\( 3 \) seconds after the initial offset and stays nearly locked on afterward. Every time the operator force changes we see that the changes are compensated quickly.

The error plots has the same performance. The initial joint-1 error of about \( 0.74\,\text{rad} \) dies out by \( t \approx 2\,\text{s} \). After that, both joints stay very close to zero, with errors between force steps hovering around \( 10^{-3}\,\text{rad} \). 

\begin{figure}[H]
    \centering
    \includegraphics[width=0.8\linewidth]{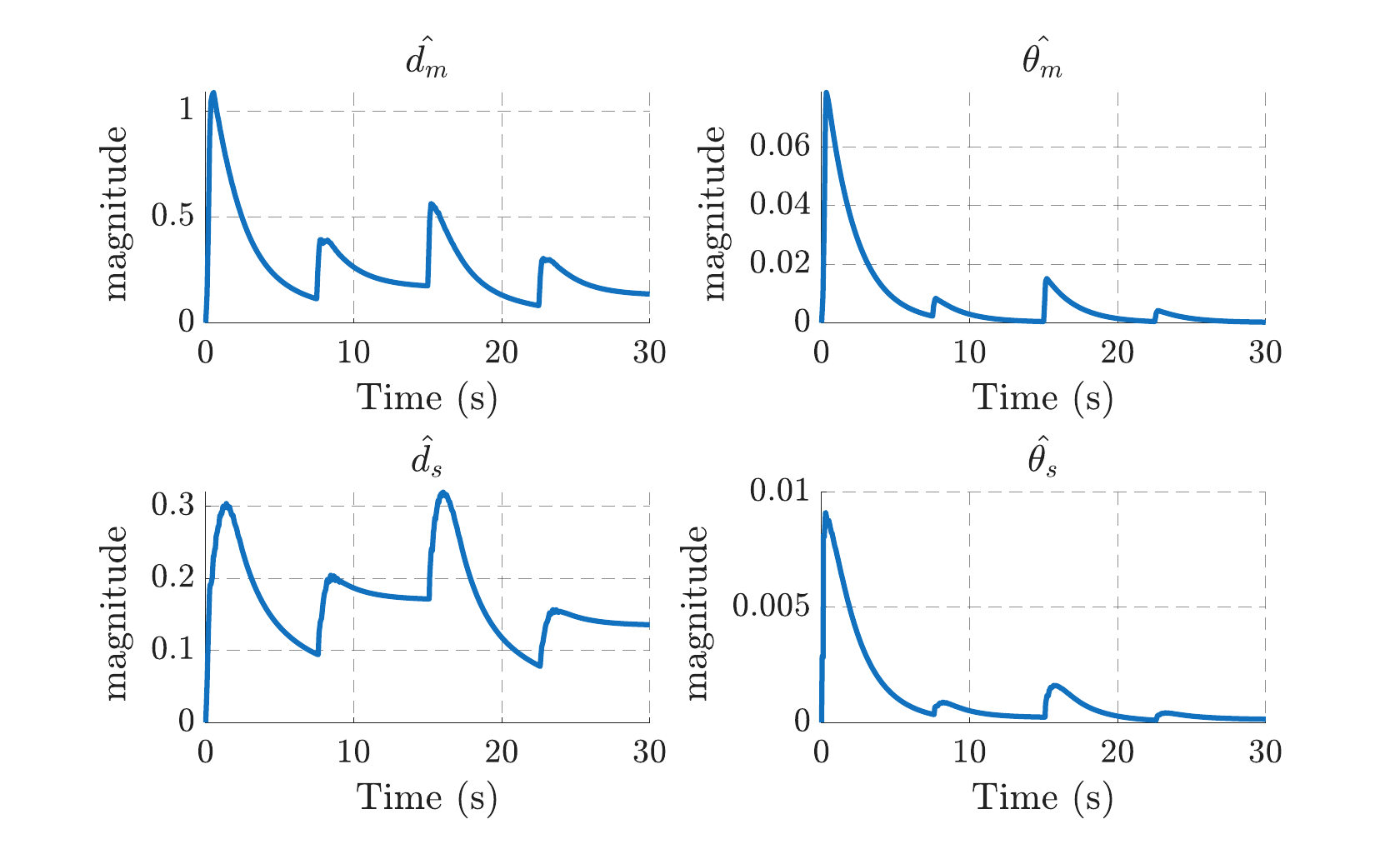}
    \caption{Master and slave estimated parameters}
    \label{fig:params}
\end{figure}

Figure (\ref{fig:params}) plots the estimated parameters for both the master and slave over time. On the master side, the initial estimates. The neural network quickly adapts to the unknown dynamics. Once the tracking error settles down, the leakage in the adaptive laws gradually pulls the estimates back. The force steps cause only minor, short-lived spikes. Importantly, the estimates never diverge; they remain well-behaved throughout.

\begin{figure}[H]
    \centering
    \includegraphics[width=0.8\linewidth]{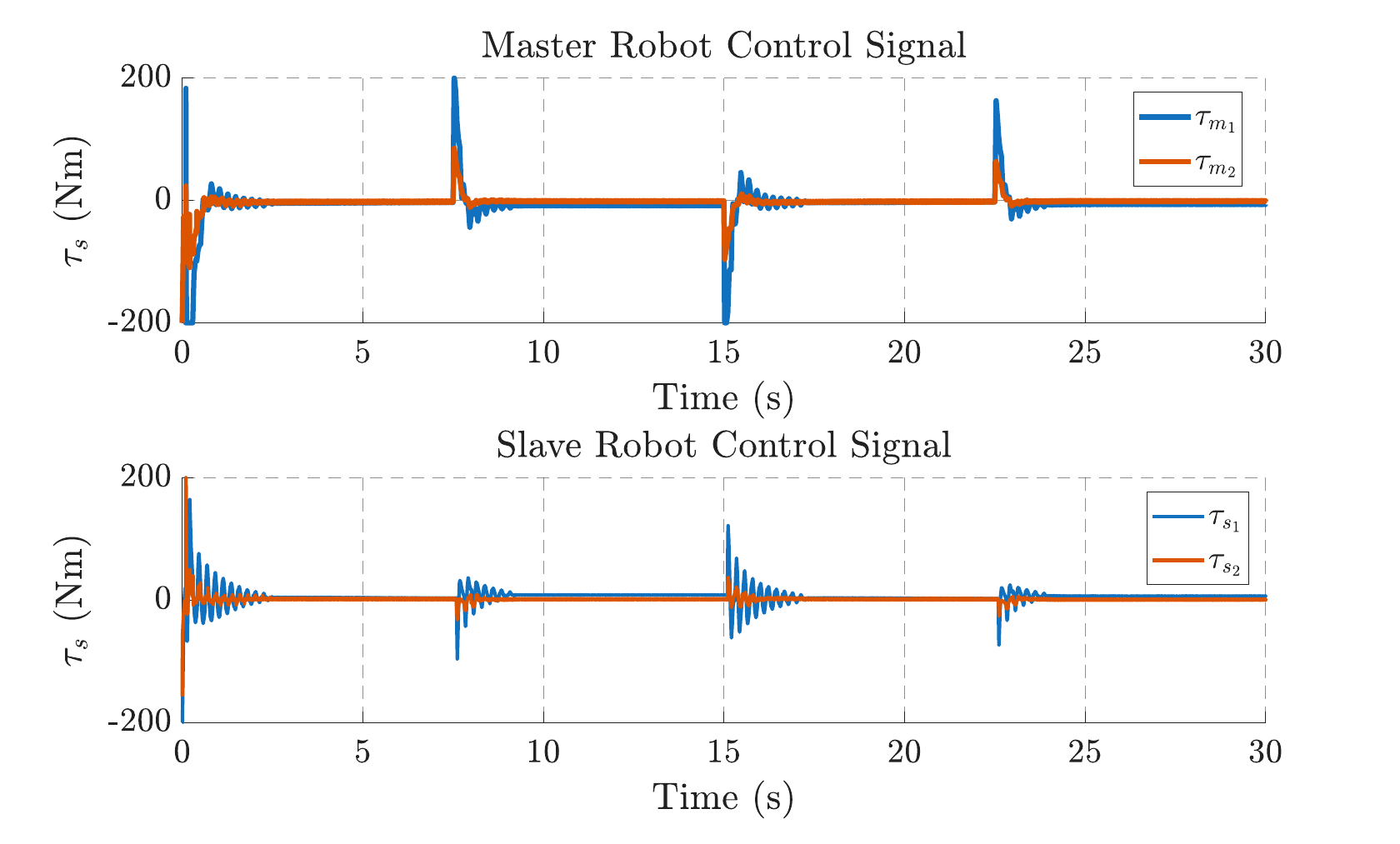}
    \caption{Control Signals of the bilateral system}
    \label{fig:cont}
\end{figure}

Figure (\ref{fig:cont}) shows the control torques. They stay within the ±200 N·m limits, only hitting and only reaching this limit in the initial startup and at the two larger force steps at \( t = 12 \) and \( 18\,\text{s} \). Most of the time, the torques are relatively small.

Some high-frequency chatter is visible, especially on the slave during the first few seconds and right after each force change. This happens when the error is large and the sliding-mode term is doing most of the work. Once the error settles into the boundary layer, the chatter fades and the torque becomes smooth.

This is the usual sliding-mode trade-off. the same switching action that gives robustness to friction and disturbances also introduces chatter while the error is still significant.

For the last part we changed the gains and kept everything else fixed, just to see how much they matter. Three settings of the sliding gain \(k\) and the robust gain \(a\) were tried, and they are listed in Table \ref{tab:gainsets}.

\begin{table}[H]
\centering
\caption{Gain settings used in the parameter study}
\label{tab:gainsets}
\begin{tabular}{c|c|c}
\hline
\textbf{Case} & \(k_m = k_s\) & \(a_m = a_s\) \\
\hline
Low gains  & \(\operatorname{diag}(30, 30)\)   & 2.5 \\
Baseline   & \(\operatorname{diag}(100, 100)\) & 5   \\
High gains & \(\operatorname{diag}(300, 300)\) & 10  \\
\hline
\end{tabular}
\end{table}

\begin{figure}[H]
    \centering
    \includegraphics[width=0.8\linewidth]{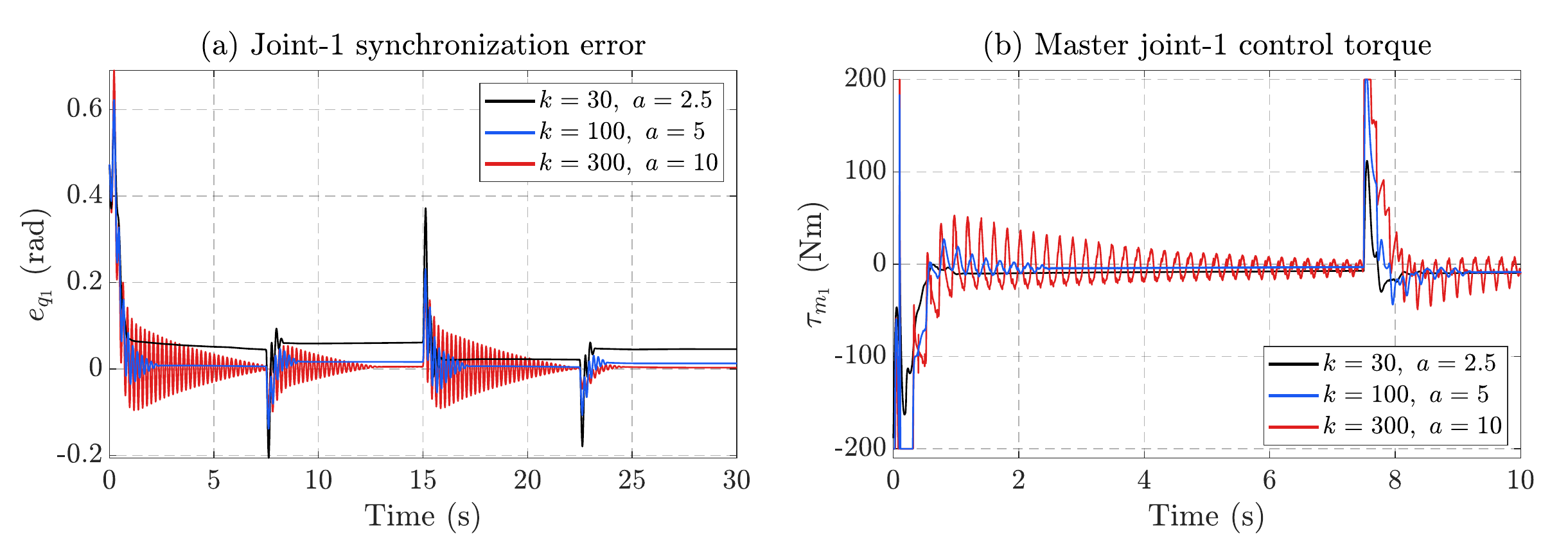}
    \caption{Effect of the controller gains: (a) joint-1 synchronization error, and (b) master joint-1 torque over the first ten seconds, for the three settings in Table \ref{tab:gainsets}.}
    \label{fig:gainstudy}
\end{figure}

Figure (\ref{fig:gainstudy}) shows the joint-1 error and the master torque for the three cases, and Table \ref{tab:gainresults} has the numbers. With the low gains the slave is clearly slower, it needs almost six seconds to settle and sits at a steady error of about \(0.05\,\text{rad}\). The good side is that the torque stays gentle and never even reaches the saturation limit. The baseline gains do best, they bring the error down in about a second and keep it the smallest of the three. Turning the gains up further does not really help. The response actually gets a bit slower again, the joint-1 error is slightly worse, and mostly what grows is the chatter and the control effort, with the root-mean-square (RMS) torque going from about 18 to 30 to 37 N·m across the three cases.

\begin{table}[H]
\centering
\caption{Performance under the three gain settings. Settling is the time for \(|e_1|\) to drop below \(0.05\,\text{rad}\).}
\label{tab:gainresults}
\begin{tabular}{c|c|c|c|c}
\hline
\textbf{Case} & Mean \(|e_1|\) (rad) & Settling (s) & RMS \(\tau_{m1}\) (N·m) & Peak \(\tau_{m1}\) (N·m) \\
\hline
Low gains  & 0.047 & 5.8 & 18 & 189 \\
Baseline   & 0.015 & 1.1 & 30 & 200 \\
High gains & 0.018 & 3.8 & 37 & 200 \\
\hline
\end{tabular}
\end{table}

So it is not really a case of higher always being better. Since the actuator saturates at ±200 N·m, pushing the gains up mainly turns into chatter and wasted effort rather than faster tracking. The baseline sits nicely between a sluggish response and an over-aggressive one, which is why we used it for the rest of the results.

\section{Conclusion}
\label{sec:conclusion}

This paper addressed the combined problem of unknown nonlinear dynamics and constant communication delay in bilateral teleoperation. The operator and environment impedances were folded into the master and slave manipulator models to obtain a combined joint-space description, and a two-sided adaptive controller was designed in which an RBF-basis-weighted feedback term is driven by a single scalar estimate of the ideal weight norm, and a second scalar estimate bounds the aggregate of joint friction, approximation residual, and external disturbance. Because both adaptive quantities are scalars, the number of parameters updated online is two per side and does not grow with the number of RBF nodes, which is the property distinguishing this construction from designs that adapt full weight matrices.

The stability analysis placed the adaptive mechanism and the communication delay inside a single Lyapunov–Krasovskii functional rather than treating the delay through a separate passivity transformation, predictor, or delay observer. Free-weighting matrices generated from the sliding-surface identities were used to obtain delay-dependent matrix conditions, under which the sliding variables, synchronization errors, and adaptive estimates were shown to be uniformly ultimately bounded, with the ultimate bound expressed explicitly in terms of the adaptation gains, leakage coefficients, and design constants.

Simulations of two 2-DOF revolute manipulators subject to joint friction, external disturbance, and stepwise operator forcing showed the slave synchronizing with the master within approximately 2 s of the initial offset and recovering rapidly after each force step, with the adaptive estimates remaining bounded throughout. A gain study over three settings indicated that raising the sliding and robust gains beyond the baseline did not improve tracking accuracy but increased the RMS control effort from approximately 18 to 37 N·m and amplified chattering, reflecting the usual switching-robustness trade-off.

Several limitations bound these conclusions. The analysis assumes constant, known-bounded delays and does not cover time-varying delay, channel asymmetry, or packet loss. The RBF approximation property holds on a compact input set, and the present argument does not establish forward invariance of that set, so the result is local in this respect. Actuator saturation is active in the simulation but is not represented in the stability analysis. The evaluation is confined to simulation of a single 2-DOF pair without measurement noise. Extending the analysis to time-varying delay, incorporating input saturation, and experimental validation on a physical master–slave pair are the natural next steps.

Another direction is to improve the convergence speed of the proposed controller. Finite- and fixed-time neural controllers have already been studied for uncertain teleoperation systems with communication delays \cite{ID5,ID13}, and similar ideas could be incorporated into the present framework to obtain a convergence-time guarantee instead of UUB alone. Reinforcement learning (RL) also offers a different extension. Actor--Critic control with RISE compensation \cite{ID2} and integral reinforcement learning with robust integral of the sign of the error (RISE) \cite{ID8} have been used to learn control policies for uncertain teleoperators with variable delays. A more recent approach combines Actor--Critic learning with RBF-based uncertainty compensation \cite{ID6}, which is particularly relevant to the RBFNN structure considered here. These methods suggest that RL could be used as part of the controller itself, for example to improve tracking performance or optimize control effort while the adaptive RBFNN continues to compensate the uncertain dynamics. A separate and more direct extension is gain tuning. The controller and adaptation gains in the present work are fixed, whereas the deep reinforcement learning-based method in \cite{ID106} provides a relevant example of adapting control gains in teleoperation systems with uncertain dynamics and communication delays. Such an approach could be used to replace the fixed feedback, robust, and adaptation gains employed in the present study.


\nocite{*}
\bibliographystyle{elsarticle-num-names}
\bibliography{related_works_v4}

\end{document}